\documentclass[a4paper, 12pt]{article}

\usepackage[sort&compress]{natbib}
\bibpunct{(}{)}{;}{a}{}{,} 

\usepackage{amsthm, amsmath, amssymb, mathrsfs, multirow, url, subfigure}
\usepackage{graphicx} 
\usepackage{ifthen} 
\usepackage{amsfonts}
\usepackage[usenames]{color}
\usepackage{fullpage}
\usepackage[shortlabels]{enumitem}
\usepackage[ruled,vlined]{algorithm2e}

\theoremstyle{plain}

\theoremstyle{definition}

\theoremstyle{remark} 

\newtheorem{ex}{Example}

\newcommand{\prob}{\mathsf{P}}

\newcommand{\pois}{{\sf Pois}}
\newcommand{\bin}{{\sf Bin}}
\newcommand{\ber}{{\sf Ber}}
\newcommand{\unif}{{\sf Unif}}
\newcommand{\nm}{{\sf N}}

\newcommand{\gam}{{\sf Gamma}}

\newcommand{\chisq}{{\sf ChiSq}}

\newcommand{\RR}{\mathbb{R}}

\newcommand{\TT}{\mathbb{T}}

\newcommand{\eps}{\varepsilon}

\newcommand{\iid}{\overset{\text{\tiny iid}}{\,\sim\,}}

\newcommand{\prior}{\mathsf{Q}}
\newcommand{\credal}{\mathscr{Q}}

\newcommand{\lPi}{\underline{\Pi}}
\newcommand{\uPi}{\overline{\Pi}}

\title{Computationally efficient variational-like approximations of possibilistic inferential models\footnote{This is an extended version of the paper \citep{cella.martin.belief2024} presented at the {\em 8th International Conference on Belief Functions}, in Belfast, UK.}}
\author{Leonardo Cella\footnote{Department of Statistical Sciences, Wake Forest University, {\tt cellal@wfu.edu}} \; and \; Ryan Martin\footnote{Department of Statistics, North Carolina State University, {\tt rgmarti3@ncsu.edu}}}
\date{\today}

\begin{document}

\maketitle 
\begin{abstract}
Inferential models (IMs) offer provably reliable, data-driven, possibilistic statistical inference.  But despite the IM framework's theoretical and foundational advantages, efficient computation is a challenge.  This paper presents a simple yet powerful numerical strategy for approximating the IM's possibility contour, or at least its $\alpha$-cut for a specified $\alpha \in (0,1)$. Our proposal starts with the specification of a parametric family that, in a certain sense, approximately covers the credal set associated with the IM's possibility measure.  Akin to variational inference, we then propose to tune the parameters of that parametric family so that its $100(1-\alpha)\%$ credible set roughly matches the IM contour's $\alpha$-cut.  This parametric $\alpha$-cut matching strategy implies a full approximation to the IM's possibility contour at a fraction of the computational cost associated with previous strategies.  

\smallskip

\end{abstract}

\section{Introduction}
\label{S:intro}

For a long time, despite Bayesians' foundational advantages, few statisticians were actually using Bayesian methods---the computational burden was simply too high.  Things changed significantly when Monte Carlo methods brought Bayesian solutions within reach.  Things changed again more recently with the advances in various approximate Bayesian computational methods, in particular, the variational approximations in \citet{blei.etal.vb.2017}, \citet{vb.tutorial.2021}, and the references therein.  The once clear lines between what was computationally feasible for Bayesians and for non-Bayesians have now been blurred, reinvigorating Bayesians' efforts in modern applications.  Dennis Lindley predicted that ``[statisticians] will all be Bayesians in 2020'' \citep{smith.convo.1995}---his prediction did not come true, but the Bayesian community is arguably stronger now than ever. 

While Bayesian and frequentist are currently the two mainstream schools of thought in statistical inference, these are not the only perspectives.  For example, Dempster--Shafer theory originated as an improvement to and generalization of both Bayesian inference and Fisher's fiducial argument.  Of particular interest to us here are the recent advances in {\em inferential models} \citep[IMs,][]{imbasics, imbook, imchar, martin.partial2}, a framework that offers Bayesian-like, data-dependent, possibilistic quantification of uncertainty about unknowns but with built-in, frequentist-like reliability guarantees.  IMs and other new/non-traditional frameworks are currently facing the same computational challenges that Bayesians faced years ago.  That is, we know what we want to compute and why, but we are currently lacking the tools to do so efficiently.  While traditional Monte Carlo methods are still useful (see Section~\ref{S:background}), the imprecision that is key to the IM's reliability guarantees implies that Monte Carlo methods alone are not enough.  In contrast to Lindley's, for Efron's prediction about fiducial-like methods---``Maybe Fisher’s biggest blunder will become a big hit in the 21st century!'' \citep{efron1998}---to come true, imprecision-accommodating advances in Monte Carlo computations are imperative.  The present paper's contribution is in this general direction, offering a new IM approximation and an efficient method for computing that approximation.

We start with a relatively simple idea that leads to a general tool for computationally efficient and statistically reliable possibilistic inference.  For reasons summarized in Section~\ref{S:background}, our focus is on the possibility-theoretic brand of IMs, which are fully determined by the corresponding contour function or, equivalently, the contour function's so-called $\alpha$-cuts.  We leverage a well-known characterization of a possibility measure's credal set as the collection of probability measures that assign probability at least $1-\alpha$ to the aforementioned $\alpha$-cuts of the contour function.  In the IM context, the $\alpha$-cuts are $100(1-\alpha)$\% confidence regions and, therefore, the contents of the IM's credal set can be justifiably interpreted as ``confidence distributions'' \citep[e.g.,][]{xie.singh.2012, schweder.hjort.2002}.  Specifically, the ``most diffuse'' element of the credal set, or ``inner probabilistic approximation,'' would assign probabilities as close to $1-\alpha$ as possible to each $\alpha$-cut.  If we could approximate this distinguished member of the IM's credal set, via Monte Carlo or otherwise, then we would be well on our way to carrying out most---if not all---of the relevant IM computations.  The challenge is that, except for very special classes of problems \citep{martin.isipta2023}, this ``inner probabilistic approximation'' would be rather complex.  We can, however, obtain a relatively simple approximation if we only require an accurate approximation of, say, a single $\alpha$-cut.  Towards this, we propose to introduce a simple parametric family of probability distributions, e.g., Gaussian, on the parameter space, with parameters partially depending on data, and then choosing those unspecified features of the parameter in such a way that the distribution (approximately) assigns probability $1-\alpha$ to a specified $\alpha$-cut, say, $\alpha=0.1$.  This is akin to variational Bayes in the sense that we propose to approximate a complex probability distribution---in our case it is the ``inner probabilistic approximation'' of the IM's possibility measure instead of a Bayesian posterior distribution---by an appropriately chosen member of a posited class of relatively simple probability distributions.  The specific, technical details our proposal here are inspired by the recent work in \citet{calibrated.boostrap} and the seemingly unrelated developments in \citet{syring.martin.scaling}.  


The remainder of the paper is organized as follows.  After some background on possibilistic IMs and their properties in Section~\ref{S:background}, we present our basic but very general variational-like IM approximation in Section~\ref{S:basic}, which relies on a combination of Monte Carlo and stochastic approximation to tune the index of the posited parametric family of approximations.  This is ideally suited for statistical inference problems involving low-dimensional parameters, and we offer a few such illustrations.
A more sophisticated version of the proposed variational-like approximation is presented in Section~\ref{S:beyond}, which is more appropriate when the problem at hand is of higher dimension, but it is suited primarily for Gaussian approximation families---which is no practical restriction.  With a more structured approximation, we can lower the computational cost by reducing the number of Monte Carlo evaluations of the IM contour.  This is illustrated in several examples, including a relatively high-dimensional problem with lasso penalization and in problems involving nuisance parameters in parametric, nonparametric, and semiparametric models, respectively.   
We emphasize that this paper is not about a new IM construction. Our contribution is a new and computationally efficient approximation to the IM solution that has been developed and studied elsewhere in the literature.  So, rather than comparing our IM approximation against other methods (e.g., Bayesian), the illustrations presented here focus on the quality of our newly proposed approximation and showing that it closely matches the often substantially more expensive exact IM solution that has already been studied.  If the proposed approximation is accurate, then the comparisons between the IM solution and other methods---see, e.g., \citet{imcens, immeta, plausfn, martin.partial2, martin.partial3, cella2024, cella.martin.imrisk, imconformal.supervised}---would be the same whether using the old and expensive computational strategy or the new and more efficient one proposed here.  
We do, however, include a brief comparison between IMs and the Bayesian approach; see Example~\ref{ex:poisson}. 
The paper concludes in Section~\ref{S:discuss} with a concise summary and a discussion of some practically relevant extensions.

\section{Background on possibilistic IMs}
\label{S:background}


The first presentation of the IM framework \citep[e.g.,][]{imbasics, imbook} relied heavily on (nested) random sets and their corresponding belief functions.  The more recent IM formulation in \citet{martin.partial2}, building off of developments in \citet{plausfn, gim}, defines the IM's possibility contour using a probability-to-possibility transform applied to the relative likelihood.  This small-but-important shift has principled motivations but we can only barely touch on this here.  The present review focuses on the possibilistic IM formulation, its key properties, and the available computational strategies. 

Consider a parametric statistical model $\{ \prob_\theta: \theta \in \TT\}$ indexed by a parameter space $\TT \subseteq \RR^d$.  Examples include $\prob_\theta = \ber(\theta)$, $\prob_\theta=\nm(\theta,1)$, $\prob_\theta = \gam(\alpha,\beta)$ with $\theta=(\alpha,\beta)$, and many others; see Section~\ref{S:illustrations}.  Nonparametric problems---see \citet{cella2024}, \citet{cella.martin.imrisk}, and \citet[][Sec.~5]{martin.partial3}---can also be handled, but we postpone discussion of this until Section~\ref{S:extensions}.  Suppose that the observable data $X^n=(X_1,\ldots,X_n)$ consists of iid samples from the distribution $\prob_\Theta$, where $\Theta \in \TT$ is the unknown/uncertain ``true value.''
The model and observed data $X^n=x^n$ together determine a likelihood function $\theta \mapsto L_{x^n}(\theta)$ and a corresponding relative likelihood function
\[ 
R(x^n,\theta) = \frac{L_{x^n}(\theta)}{\sup_\vartheta L_{x^n}(\vartheta)}.
\]
Throughout we will implicitly assume that the denominator is finite. As a running example, consider a binomial problem where $X^n$ represent the outcomes of $n$ independent Bernoulli trials with unknown probability of success $\Theta$. Then the relative likelihood is
\begin{equation}\label{eq:binomRL}
R(x^n,\theta) = \left( \frac{n\theta}{\sum_{i=1}^n x_i} \right)^{\sum_{i=1}^n x_i} \left( \frac{n-n\theta}{n-\sum_{i=1}^n x_i} \right)^{n-\sum_{i=1}^n x_i}, \quad \theta \in [0,1].
\end{equation}

The relative likelihood, $\theta \mapsto R(x^n, \theta)$, itself is a data-dependent possibility contour, i.e., it is a non-negative function such that $\sup_\theta R(x^n,\theta) = 1$ for almost all $x^n$.  This contour, in turn, determines a possibility measure, or consonant plausibility function, that can be used for uncertainty quantification about $\Theta$, given $X^n=x^n$, which has been extensively studied \citep[e.g.,][]{shafer1982, wasserman1990b, denoeux2006, denoeux2014}.  Specifically, uncertainty quantification here would be based on the possibility assignments
\[ H \mapsto \sup_{\theta \in H} R(x^n, \theta), \quad H \subseteq \TT, \]
where $H$ represents a hypothesis about $\Theta$.  This purely likelihood-driven possibility measure has a number of desirable properties: for example, inference based on thresholding the possibility scores (with model-independent thresholds) satisfies the likelihood principle \citep[e.g.,][]{birnbaum1962, basu1975}, and it is asymptotically consistent (under standard regularity conditions) in the sense that it converges in $\prob_\Theta$-probability to a possibility measure concentrated on $\Theta$ as $n \to \infty$.  
What the purely likelihood-based possibility measure above lacks, however, is a calibration property (relative to the posited model) that gives meaning or belief-forming inferential weight to the ``possibility'' it assigns to hypotheses about the unknown $\Theta$.  More specifically, if we offer a possibility measure as our quantification of statistical uncertainty, then its corresponding credal set should contain probability distributions that are statistically meaningful.  We are assuming vacuous prior information, so there are no meaningful or distinguished Bayesian posterior distributions.  The only other natural property to enforce on the credal set's elements is that they be ``confidence distributions.''  But, as in \eqref{eq:credal.char}, this interpretation requires the $\alpha$-cuts of the relative likelihood, i.e., $\{\theta: R(x^n, \theta) > \alpha\}$, to be $100(1-\alpha)$\% confidence sets for $\Theta$, which generally they are not.  Therefore, the relative likelihood alone is not enough. 

Fortunately, it is at least conceptually straightforward to achieve this calibration by applying what \citet{martin.partial} calls ``validification''---a version of the probability-to-possibility transform \citep[e.g.,][]{dubois.etal.2004, hose2022thesis}.  In particular, for observed data $X^n=x^n$, the possibilistic IM's contour is defined as
\begin{equation}
\label{eq:contour}
\pi_{x^n}(\theta) = \prob_\theta\bigl\{ R(X^n,\theta) \leq R(x^n, \theta) \bigr\}, \quad \theta \in \TT,
\end{equation}
and the possibility measure---or upper probability---is 
\[ \uPi_{x^n}(H) = \sup_{\theta \in H} \pi_{x^n}(\theta), \quad H \subseteq \TT. \]
A corresponding necessity measure, or lower probability, is defined via conjugacy as $\lPi_{x^n}(H) = 1 - \uPi_{x^n}(H^c)$. This measure plays a crucial role in quantifying the support for hypotheses of interest \citep{cella.martin.probing}; see Example~\ref{ex:poisson}. For our binomial running example with relative likelihood \eqref{eq:binomRL}, the possibility contour based on $X^n=x^n$ is
\begin{equation}
\label{eq:binom.contour}
\pi_{x^n}(\theta) = \sum_{s=0}^n 1\{R(s,\theta) \leq R(x^n,\theta)\} \, p_\theta(s), 
\end{equation}
where $1(\cdot)$ is the indicator function and $p_\theta$ is the probability mass function of $\bin(n,\theta)$. This can be easily evaluated, without Monte Carlo, and it is displayed in Figure~\ref{fig:two.examples.a}(a).

An essential feature of above IM construction is its so-called {\em validity property}:
\begin{equation}
\label{eq:valid}
\sup_{\Theta \in \TT} \prob_\Theta\bigl\{ \pi_{X^n}(\Theta) \leq \alpha \bigr\} \leq \alpha, \quad \text{for all $\alpha \in [0,1]$}. 
\end{equation}
The proof of \eqref{eq:valid} is a simple application of the now-standard probability integral transform \citep[originally due to][Sec.~21.1]{fisher1973b}, i.e., if a random variable $Z$ has distribution function $G$, then the random variable $G(Z)$ is stochastically no larger than $\unif(0,1)$.  The property \eqref{eq:valid} has a number of important consequences.  First, \eqref{eq:valid} is exactly the calibration property required to ensure a ``confidence'' connection that the relative likelihood on its own misses.  In particular, \eqref{eq:valid} immediately implies that the set 
\begin{equation}
\label{eq:conf.set}
C_\alpha(x^n) = \{\theta \in \TT: \pi_{x^n}(\theta) > \alpha\}, 
\end{equation}
indexed by $\alpha \in [0,1]$, is a $100(1-\alpha)$\% frequentist confidence set in the sense that its coverage probability is at least $1-\alpha$:  
\[ \sup_{\Theta \in \TT} \prob_\Theta\bigl\{ C_\alpha(X^n) \not\ni \Theta \bigr\} \leq \alpha, \quad \alpha \in [0,1]. \]
Moreover, it readily follows that 
\begin{equation}
\label{eq:valid.alt}
\sup_{\Theta \in H} \prob_\Theta\bigl\{ \uPi_{X^n}(H) \leq \alpha \bigr\} \leq \alpha, \quad \text{all $\alpha \in [0,1]$, all $H \subseteq \TT$}. 
\end{equation}
In words, the IM is valid, or calibrated, if it assigns possibility $\leq \alpha$ to true hypotheses at rate $\leq \alpha$ as a function of data $X^n$.  This is precisely where the IM's aforementioned ``inferential weight'' comes from: \eqref{eq:valid.alt} implies that we do not expect small $\uPi_{x^n}(H)$ when $H$ is true, so we are inclined to doubt the truthfulness of those hypotheses $H$ for which $\uPi_{x^n}(H)$ is small.  In addition, the above property ensures that the possibilistic IM is safe from false confidence \citep{balch.martin.ferson.2017, martin.nonadditive, martin.belief2024}, unlike all default-prior Bayes and fiducial solutions.  An even stronger, {\em uniform-in-hypotheses} version of \eqref{eq:valid.alt} holds, as shown/discussed in \citet{cella.martin.probing}:
\[ \sup_{\Theta \in \TT} \prob_\Theta\bigl\{ \uPi_{X^n}(H) \leq \alpha \text{ for some $H$ with $H \ni \Theta$} \bigr\} \leq \alpha, \quad \text{all $\alpha \in [0,1]$}. \]
The ``for some $H$ with $H \ni \Theta$'' event can be seen as a union of every hypothesis $H$ that contains $\Theta$, which makes it a much broader event than that associated with any single fixed $H$ in \eqref{eq:valid.alt}. Consequently, regardless of the number of hypotheses evaluated or the manner in which they are selected---even if they are data-dependent---the probability that even a single suggestion from the IM is misleading remains controlled at the specified level.
For further details about and discussion of the possibilistic IM's properties, its connection to Bayesian/fiducial inference, etc, see \citet{martin.partial2, martin.partial3, martin.isipta2023}. 

In a Bayesian analysis, inference is based on summaries of the data-dependent posterior distribution, e.g., posterior probabilities of scientifically relevant hypotheses, expectations of loss/utility functions, etc.  And all of these summaries boil down to integration involving the probability density function that determines the posterior.  Virtually the same statement is true of the possibilistic IM: the lower--upper probability pairs for scientific relevant hypotheses, lower--upper expectations of loss/utility functions, etc.~boil down to optimization involving the possibility contour $\pi_{x^n}$.  For example, if the focus is on a certain feature $\Phi = g(\Theta)$ of $\Theta$, then the Bayesian can find the marginal posterior distribution for $\Phi$ by integrating the posterior density for $\Theta$.  Similarly, the corresponding marginal possibilistic IM has a contour that is obtained by optimizing $\pi_{x^n}$:
\[ \pi_{x^n}^\text{marg}(\phi) = \sup_{\theta:g(\theta) = \phi}\pi_{x^n}(\theta), \quad \phi \in g(\TT). \]    
Importantly, unlike with Bayesian integration, the IM's optimization operation ensures that the validity property inherent in $\pi_{x^n}$ is transferred to $\pi_{x^n}^\text{marg}$, which implies that the IM's marginal inference about $\Phi$ is safe from false confidence.  

The above-described operations applied to the IM's contour function \eqref{eq:contour}, to obtain upper probabilities or to eliminate nuisance parameters, are all special cases of Choquet integrals \citep[e.g.][App.~C]{lower.previsions.book}.  These more general Choquet integrals can be statistically relevant in formal decision-making contexts \citep{imdec}, etc.  That is, if $\ell_a(\Theta)$ represents the loss associated with taking action $a$ when the state of the world is $\Theta$, then the possibilistic IM provides an assessment of the risk associated with action $a$ that accounts for uncertainty in $\Theta$, given $x^n$, via the Choquet integral:
\[ \uPi_{x^n} \ell_a := \int_0^1 \Bigl\{ \sup_{\theta: \pi_{x^n}(\theta) > s} \ell_a(\theta) \Bigr\} \,ds. \]
Then, for example, one might choose the action $\hat a(x^n)$ that minimizes this upper expected loss. The point is that there are non-trivial operations to be carried out even {\em after} the IM's contour function has been obtained provides genuine, practical motivation for us to find the simplest and most efficient contour approximation as possible.  

While the IM construction is conceptually simple and its properties are strong, computation can be a challenge.  The issue is that we rarely have the sampling distribution of the relative likelihood $R(X^n,\theta)$, under $\prob_\theta$, available in closed form to facilitate exact computation of $\pi_{x^n}$. 
So, instead, the go-to strategy is to approximate that sampling distribution using Monte Carlo at each value of $\theta$ on a sufficiently fine grid \citep[e.g.,][]{martin.partial2, hose.hanss.martin.belief2022}.  That is, the possibility contour is approximated as 
\begin{equation}
\label{eq:naive}
\pi_{x^n}(\theta) \approx \frac1M \sum_{m=1}^M 1\{ R(X_{m,\theta}^n, \theta) \leq R(x^n, \theta) \}, \quad \theta \in \TT, 
\end{equation}
where $X_{m,\theta}^n$ consists of $n$ iid samples from $\prob_\theta$ for $m=1,\ldots,M$.  The above computation is feasible at one or a few different $\theta$ values, but it is common that this needs to be carried out over a sufficiently fine grid covering the (relevant area) of the often multi-dimensional parameter space $\TT$.  For example, the confidence set in \eqref{eq:conf.set} requires that we can solve the equation $\pi_{x^n}(\theta)=\alpha$, or at least find which $\theta$'s satisfy $\pi_{x^n}(\theta) \geq \alpha$, and a naive approach is to compute the contour over a huge grid and then keep those that (approximately) solve the aforementioned equation.  The computational complexity associated with this is $O(M g^d)$, where $M$ is the Monte Carlo sample size, $d$ is the dimension of the parameter space, and $g$ is number of grid points in each dimension of $\TT$; this amounts to lots of wasted computations.  More generally, the relevant summaries of the IM output involve optimization of the contour, and carrying out this optimization numerically requires many contour function evaluations.  Simple tweaks to this most-naive approach can be employed in certain cases, e.g., importance sampling, but these adjustments require problem-specific considerations and they cannot be expected to offer substantial improvements in computational efficiency.  This is a serious bottleneck, so new and not-so-naive computational strategies are desperately needed. 




\section{Basic variational-like IMs}
\label{S:basic}

\subsection{Setup}

The Monte Carlo-based strategy reviewed above is not the only way to numerically approximate the possibilistic IM.  Another option is an analytical ``Gaussian'' approximation (see below) based on the available large-sample theory \citep{imbvm.ext}.  The goal here is to strike a balance between the (more-or-less) exact but expensive Monte Carlo-based approximation and the rough but cheap large-sample approximation.  To strike this balance, we choose to focus on a specific feature of the possibilistic IM's output, namely, the confidence sets $C_\alpha(x^n)$ in \eqref{eq:conf.set}, and choose the approximation such that it at least matches the given confidence set exactly.  Our specific proposal resembles the variational approximations that are now widely used in Bayesian analysis, where first a relatively simple family of candidate probability distributions is specified and then the approximation is obtained by finding the candidate that minimizes (an upper bound on) the distance/divergence from the exact posterior distribution.  Our approach differs in the sense that we are aiming to approximate a possibility measure by (the probability-to-possibility transform applied to) an appropriately chosen probability distribution. 

Following \citet{destercke.dubois.2014}, \citet{cuoso.etal.2001}, and others, the possibilistic IM's credal set, $\mathscr{C}(\uPi_{x^n})$, which is the set of precise probability distributions dominated by $\uPi_{x^n}$, has a remarkably simple and intuitive characterization:
\begin{equation}
\label{eq:credal.char}
\prior_{x^n} \in \mathscr{C}(\uPi_{x^n}) \iff \prior_{x^n}\{ C_\alpha(x^n) \} \geq 1-\alpha, \quad \text{for all $\alpha \in [0,1]$}.
\end{equation}
(Where convenient, we will replace the subscript ``$x^n$'' by a subscript ``$n$''---e.g., $\prior_n$ and $\uPi_n$ instead of $\prior_{x^n}$ and $\uPi_{x^n}$---to simplify the notation in what follows.)  That is, a data-dependent probability $\prior_n$ is consistent with $\uPi_n$ if and only if, for each $\alpha \in [0,1]$, it assigns at least $1-\alpha$ mass to the IM's confidence set $C_\alpha(x^n)$ in \eqref{eq:conf.set}.  Furthermore, the ``best'' inner probabilistic approximation of the possibilistic IM, if it exists, corresponds to a $\prior_n^\star$ such that $\prior_n^\star\{ C_\alpha(x^n) \} = 1-\alpha$ for all $\alpha \in [0,1]$.  For a certain special class of statistical models, 
\citet{martin.isipta2023} showed that this best inner approximation corresponds to Fisher's fiducial distribution and the default-prior Bayesian posterior distribution.  Beyond this special class of models, it is not clear how to find a best inner approximation.  A less ambitious goal is to find, for a fixed $\alpha$, a probability distribution $\prior_{n,\alpha}^\star$ such that 
\begin{equation}
\label{eq:objective}
\prior_{n, \alpha}^\star\{ C_\alpha(x^n) \} = 1-\alpha. 
\end{equation}
Our goal here is to develop a general strategy for finding, for a given $\alpha \in (0,1)$, a probability distribution $\prior_{n,\alpha}^\star$ that (at least approximately) solves the equation in \eqref{eq:objective}.  Once identified, we can reconstruct relevant features of the possibilistic IM via (Bayesian-like) Monte Carlo sampling from this $\prior_{n,\alpha}^\star$.

\subsection{Proposed approximation}

We propose to start with a parametric class of data-dependent probability distributions $\credal^\text{var} = \{ \prior_n^\xi: \xi \in \Xi \}$ indexed by a generic space $\Xi$.  An important example is the case where the $\prior_n^\xi$'s are Gaussian distributions with mean vector and/or covariance matrix depending on (data and) $\xi$ in some specified way.  Specifically, since the possibilistic IM contour's peak is at the maximum likelihood estimator $\hat\theta_n = \hat\theta_{x^n}$, it makes sense to fix the mean vector of the Gaussian $\prior_n^\xi$ at $\hat\theta_n$; for the covariance matrix, however, a natural choice would be $\xi^2 \, J_n^{-1}$, where $J_n = J_{x^n}$ is the observed Fisher information matrix depending on data and on the posited statistical model.  This Gaussian family is a very reasonable default choice of $\credal^\text{var}$ given that $\prior_n^\xi$ with $\xi=1$ is asymptotically the best inner probabilistic approximation to $\uPi_n$; see \citet{imbvm.ext}.  So, our proposal is to insert some additional flexibility in the Gaussian approximation by allowing the spread to expand or contract depending on whether $\xi > 1$ or $\xi < 1$.  While a Gaussian-based approximation is natural, choosing $\credal^\text{var}$ to be a Gaussian family is not the only option for $\credal^\text{var}$; see Example~\ref{ex:multinomial} below.  Indeed, if the parameter space has structure that is not present in the usual Euclidean space where the Gaussian is supported, e.g., if $\TT$ is the probability simplex, then choosing $\credal^\text{var}$ to respect that structure makes perfect sense.  


The one high-level condition imposed on $\credal^\text{var}$ is that it be sufficiently flexible, i.e., as $\xi$ varies, the $\prior_{n,\alpha}^\xi$-probability of the possibilistic IM's $\alpha$-cut takes values smaller and larger than the target level $1-\alpha$.  This clearly holds for the Gaussian approximation described above, since $\xi$ controls the spread of $\prior_{n,\alpha}^\xi$ to the extent that the aforementioned $\prior_{n,\alpha}^\xi$-probability can be made arbitrarily small or large by taking $\xi$ sufficiently small or large, respectively.  This mild condition can also be readily verified for virtually any other (reasonable) approximation family $\credal^\text{var}$.  

Given such a suitable choice of approximation family $\credal^\text{var}$, indexed by a parameter $\xi \in \Xi$, our proposed procedure is as follows.  Define an objective function 
\begin{equation}
\label{eq:fxi}
f_\alpha(\xi) = \prior_{n, \alpha}^\xi(\{ \theta: \pi_{x^n}(\theta) > \alpha \}) - (1-\alpha),    
\end{equation}
so that solving \eqref{eq:objective} boils down to finding a root of $f_\alpha$.  If the probability on the right-hand side of \eqref{eq:fxi} could be evaluated in closed-form, then one could apply any of the standard root-finding algorithms to solve this, e.g., Newton--Raphson. However, this $\prior_{n, \alpha}^\xi$-probability typically cannot be evaluated in closed-form so, instead, $f_\alpha$ can be approximated via Monte Carlo with $\hat f_\alpha$ defined as 
\begin{equation}
\label{eq:fxihat}
\hat f_\alpha(\xi) = 
\frac1K \sum_{k=1}^K 1\{ \pi_{x^n}(\Theta_k^\xi) > \alpha \} - (1-\alpha),
\end{equation}
where $\Theta_1^\xi, \ldots, \Theta_K^\xi \iid \prior_{x^n}^\xi$. Presumably, the aforementioned samples are cheap for every $\xi$ because the family $\credal$ has been specified by the user, but we do still need $M$ many Monte Carlo samples to evaluate $\pi_{x^n}(\Theta_k^\xi)$ for each $k$.  This has computational complexity $O(MK)$ so, except perhaps for low-dimensional cases, e.g., $d \in \{1,2\}$, this modification amounts to a substantial computaional savings compared to the naive strategy in \eqref{eq:naive}.  There is an additional computational step, described next, but this has bounded complexity, so the overall complexity remains at $O(MK)$.

Only having an unbiased estimator of the objective function requires some adjustments to the numerical routine.  In particular, rather than a Newton--Raphson routine that assumes the function values are noiseless, we must use a {\em stochastic approximation} algorithm \citep[e.g.,][]{syring.martin.isipta21, syring.martin.scaling, martinghosh, kushner, robbinsmonro} that is adapted to noisy function evaluations of $\hat f_\alpha$. The basic Robbins--Monro algorithm seeks the root of \eqref{eq:fxi} through the updates
\[ \xi^{(t+1)} = \xi^{(t)} \pm w_{t+1} \, \hat f_\alpha(\xi^{(t)}), \quad t\geq 0,\]
where ``$\pm$'' depends on whether $\xi \mapsto f_\alpha(\xi)$ is decreasing or increasing, $\xi^{(0)}$ is an initial guess, and $(w_t)$ is a deterministic step size sequence that satisfies
\[ \sum_{t=1}^\infty w_t = \infty \quad \text{and} \quad \sum_{t=1}^\infty w_t^2 < \infty. \]
Pseudocode for our proposed approximation is given in Algorithm~\ref{algo:varim}.  To summarize, we are proposing a simple data-dependent probability distribution for $\Theta$ whose probability-to-possibility contour closely matches the IM's contour at least at the specified threshold $\alpha$.  More specifically, a reasonable choice of probability distribution is a Gaussian, and the approach presented in Algorithm~\ref{algo:varim} scales the covariance matrix so that its corresponding contour function accurately approximates the IM's contour at threshold $\alpha$.  

\begin{algorithm}[t]
\SetAlgoLined
requires: data $x^n$ and ability to evaluate $\pi_n = \pi_{x^n}$\; 
initialize: $\alpha$-level, class $\credal^\text{var} = \{ \prior_n^\xi: \xi \in \Xi \}$, guess $\xi^{(0)}$, step size sequence $(w_t)$, Monte Carlo sample size $M$, and convergence threshold $\eps > 0$\; 
set: {\tt stop = FALSE}, $t=0$\; 
\While{{\tt !stop}}{
sample $\Theta_1, \ldots, \Theta_M \iid \prior_{n}^{\xi^{(t)}}$\;
evaluate $\hat f_\alpha(\xi^{(t)})$ as in \eqref{eq:fxihat} using $(\Theta_1, \ldots, \Theta_M)$\;
update $\xi^{(t+1)} = \xi^{(t)} \pm w_{t+1} \, \hat{f}_\alpha(\xi^{(t)})$\;
\eIf{$|\xi^{(t+1)} - \xi^{(t)}| < \eps$}{
  $\hat\xi = \xi^{(t+1)}$\;
  {\tt stop = TRUE}\;
}{
  $t \gets t+1$\;
}
}
return $\hat\xi$\; 
\caption{$\alpha$-specific variational approximation of the IM}
\label{algo:varim}
\end{algorithm}
 
Under certain mild conditions, the sequence $(\xi^{(t)}: t \geq 0)$ as defined above converges in probability to the root of $f_\alpha$ in \eqref{eq:fxi}. If $\hat\xi$ is the value returned when the algorithm reaches practical convergence, e.g., when $|\hat f_\alpha(\xi^{(t)})|$ or the change $|\xi^{(t+1)}-\xi^{(t)}|$ is smaller than some specified threshold, then we set $\widehat\prior_{n,\alpha} = \prior_{n,\alpha}^{\hat\xi}$.  This distribution should be a reasonably accurate approximation of the IM possibility measure's inner approximation, i.e., the ``most diffuse'' member of $\mathscr{C}(\uPi_{x^n})$.  Consequently, the probability-to-possibility transform in \eqref{eq:contour} applied to $\widehat\prior_{x^n,\alpha}$ should be a reasonably accurate approximation of the exact possibilistic IM contour $\pi_{x^n}$, at least in terms of their upper $\alpha$-cuts.  The illustrations in the following sections confirm that this and the more sophisticated approximation presented in Section~\ref{S:beyond} below are indeed reasonably accurate. 

As emphasized above, a suitable choice of $\credal^\text{var}$ depends on context. An important consideration in this choice is the ability to perform exact computations of the approximate contour determined by the inner approximation  $\widehat\prior_{n,\alpha}$.  This is the case for the aforementioned normal variational family $\credal^\text{var}$, with mean $\hat\theta_{n}$ and covariance $\hat \xi^2 J_{n}^{-1}$, as
\begin{equation}\label{eq:normalappox}
\hat \pi_{x^n}(\theta) = 1 - G_d\{(\theta - \hat\theta_{n})^\top J_{n}(\theta - \hat\theta_{n}) / \hat\xi^2 \},    
\end{equation}
where $G_d$ is the $\chisq(d)$ distribution function.  This closed-form expression makes summaries of the contour, which are approximations of certain features (e.g., Choquet integrals) of the possibilistic IM $\uPi_{x^n}$, relatively easy to obtain numerically.

\subsection{Numerical illustrations}
\label{S:illustrations}

Our first goal here is to provide a proof-of-concept for the proposed approximation.  Towards this, we present a few low-dimensional illustrations where we can visualize both the exact and approximation IM contours and directly assess the quality of the approximation.  In all but Example~\ref{ex:multinomial} below, we use the normal variational family $\credal$ with mean $\hat\theta_n$ and covariance $\xi^2 J_n^{-1}$ as described above, with $\xi$ to be determined.  All of the examples display the variational-like IM approximation $\widehat\prior_{n,\alpha}$ based on $\alpha=0.1$, $M=200$ Monte Carlo samples, step sizes $w_t = 2(1+t)^{-1}$, and convergence threshold $\eps=0.005$. 

\begin{figure}[t]
\begin{center}
\subfigure[Example~1: Bernoulli]{\scalebox{0.57}{\includegraphics{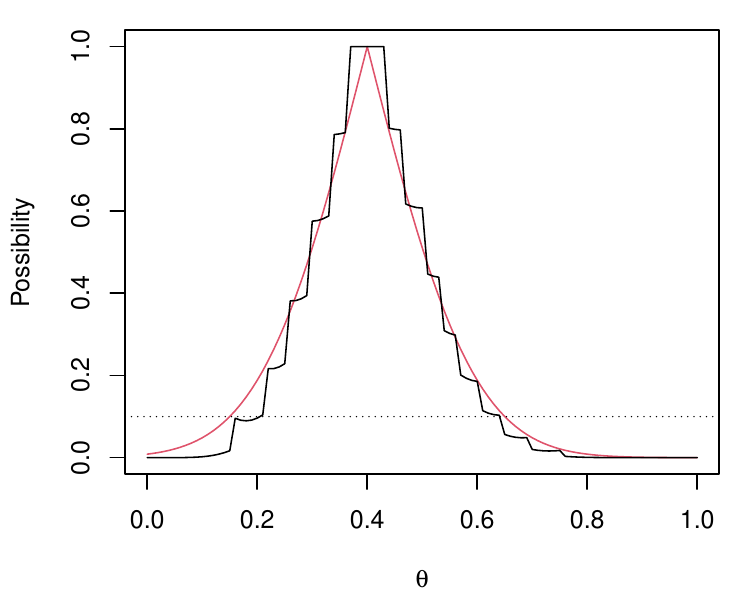}}}
\subfigure[Example~2: Bivariate normal]{\scalebox{0.57}{\includegraphics{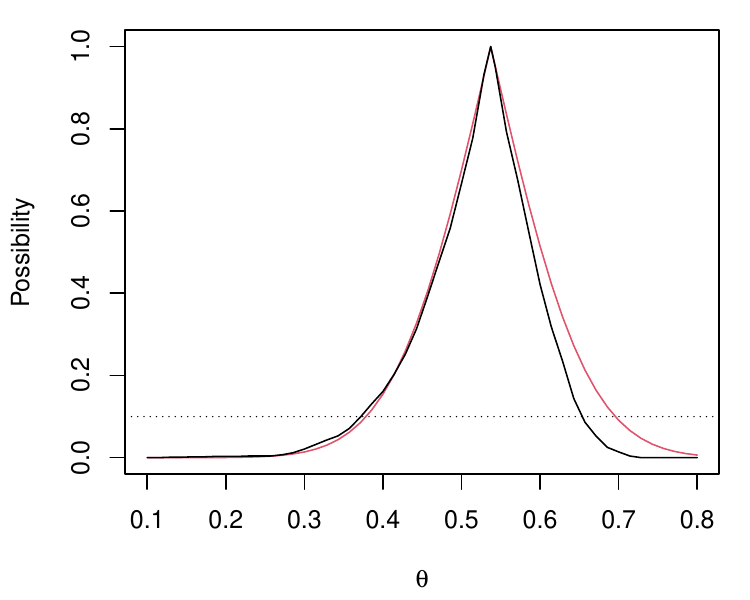}}}
\subfigure[Example~3: Logistic regression]{\scalebox{0.57}{\includegraphics{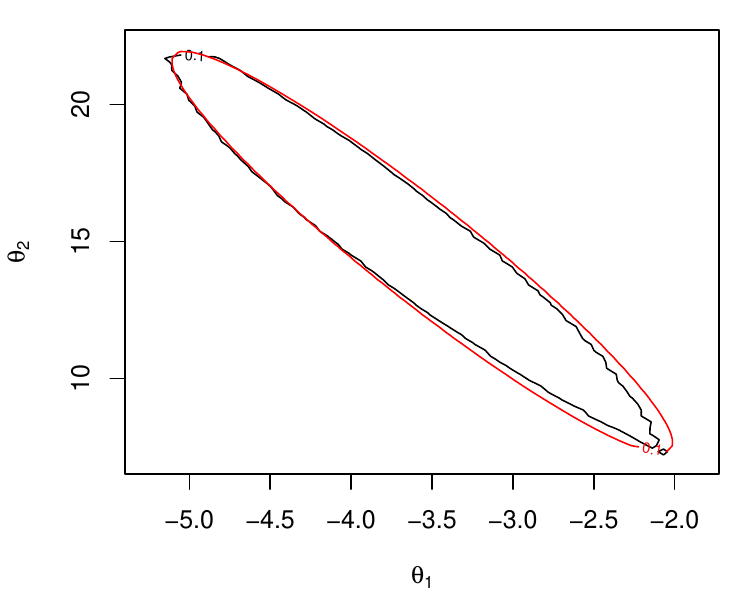}}}
\subfigure[Example~4: Multinomial]{\scalebox{0.57}{\includegraphics{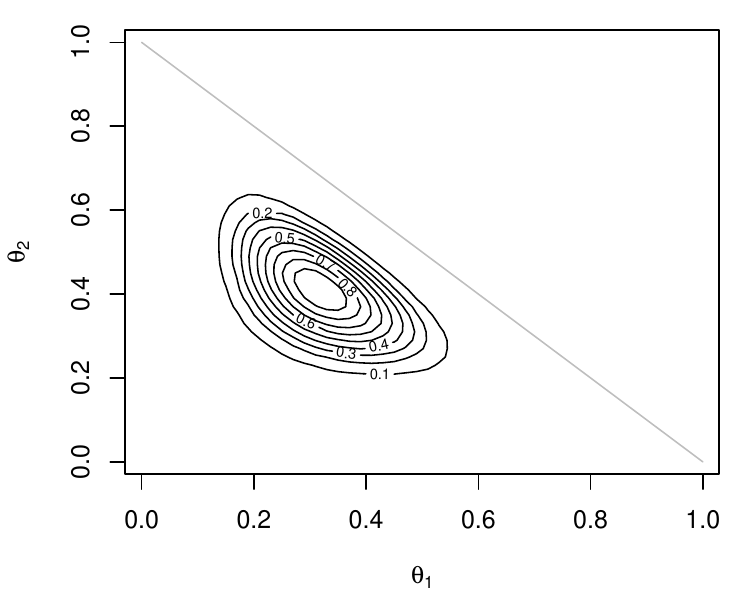}}}
\end{center}
\caption{Exact (black) and approximate (red) IM contours; 
 Panel~(d) shows only the approximate contour (black), which is fully supported on the lower triangle corresponding to the probability simplex.}
\label{fig:two.examples.a}
\end{figure}

\begin{ex} 
\label{ex:binom}
Recall the binomial example explored in Section~\ref{S:background}, where $X^n$ consists of $n$ iid $\ber(\Theta)$ random variables. An expression for the exact IM possibility contour based on $X^n=x^n$ is given in \eqref{eq:binom.contour} and this is displayed in Figure~\ref{fig:two.examples.a}(a) for $n=15$ and $x^n$ such that $\sum_{i=1}^nx_i=6$. Figure~\ref{fig:two.examples.a}(a) also shows the contour corresponding to the proposed Gaussian-based variational approximation. Note that the two contours closely agree, especially at the level $\alpha=0.1$ specifically targeted.  
\end{ex}

\begin{ex}
\label{ex:bivariate}
Suppose $X^n$ consists of iid bivariate normal pairs with zero means and unit variances, with common density function
\[ p_\theta(x) = \frac{1}{2\pi(1-\theta^2)^{1/2}} \, \exp\Bigl\{ -\frac{x_1^2 - 2 \theta x_1 x_2 + x_2^2}{2(1-\theta^2)} \Bigr\}, \quad x=(x_1,x_2), \quad \theta \in [-1,1]. \]
Inference on the unknown correlation $\Theta$ is a surprisingly challenging problem \citep[e.g.,][]{basu1964, reid2003, imcond}.  Indeed, despite being a scalar exponential family model, it does not have a one-dimensional sufficient statistic and, moreover, there are a variety of different ancillary statistics on which to condition, leading to different solutions.  
Figure~\ref{fig:two.examples.a}(b) shows the exact IM contour, based on a naive Monte Carlo implementation of \eqref{eq:contour}, and the approximation for simulated data of size $n=50$ with true correlation 0.5.  The exact contour has some asymmetry that the normal approximation cannot perfectly accommodate, but it makes up for this imperfection with a slightly wider upper 0.1-level set. A quantitative assessment of the computational efficiency and accuracy of the proposed approximation is presented in Example~\ref{ex:bivariate.again} below.
\end{ex}

\begin{ex}
\label{ex:logistic}
The data presented in Table~8.4 of \citet{ghosh-etal-book} concerns the relationship between exposure to chloracetic acid and mouse mortality. A simple logistic regression model is to be fit in order to relate the binary death indicator ($y$) with the levels of exposure ($u$) to chloracetic acid for the dataset's $n=120$ mice.  That is, $X^n$ consists of independent pairs $X_i = (U_i, Y_i)$ and a conditionally Bernoulli model for $Y_i$, given $U_i$, with mass function 
\[ p_\theta(y \mid u) = F(\theta_1 + \theta_2 u)^y \, \{1 - F(\theta_1 + \theta_2 u)\}^{1-y}, \quad \theta=(\theta_1, \theta_2) \in \RR^2, \]
where $F(z) = (1 + e^{-z})^{-1}$ is the logistic distribution function.  The corresponding likelihood cannot be maximized in closed-form, but doing so numerically is routine.  The maximum likelihood estimator and the corresponding observed information matrix lead to the asymptotically valid inference reported by standard statistical software packages.  For exact inference, however, the computational burden is heavier: evaluating the validified relative likelihood over a sufficiently fine grid of $\theta$ values is rather expensive.  Figure~\ref{fig:two.examples.a}(c) presents the $0.1$-level set of the exact IM possibility contour for the regression coefficients, based on a naive Monte Carlo implementation of \eqref{eq:contour}, alongside the proposed variational approximation. The variational method is nearly 2.5 times faster than the naive one, yet the two contours align almost perfectly.
\end{ex}

\begin{ex}
\label{ex:multinomial}
Consider $n$ iid data points sampled from $\{1,2,\ldots,K\}$ with unknown probabilities $\Theta = (\Theta_1,\ldots,\Theta_K)$ belong the probability simplex $\TT$ in $\RR^K$.  The frequency table, $X = (X_1,\ldots,X_K)$, with $X_k$ being the frequency of category $k$ in the sample of size $n$, is a sufficient statistic, having a multinomial distribution with parameters $(n,K,\Theta)$; the probability mass function is 
\[ p_\theta(x) = \frac{n!}{\prod_{k=1}^n x_k!} \prod_{k=1}^K \theta_k^{x_k}, \quad \theta \in \TT.  \]
Here we use a Dirichlet variational family $\prior_{x^n}^\xi$, parametrized in terms of the mean vector $m \in \TT$ and precision $s > 0$, with density function 
\[ q(\theta) \propto \prod_{k=1}^K \theta_k^{s m_k-1}, \quad \theta \in \TT, \]
where, naturally, we take the mean to be the maximum likelihood estimator, $\hat\theta=n^{-1}X$, and the precision to be $n\xi$, where $\xi > 0$ is to be determined. Of course, a Gaussian variational approximation could also be used here, but our use of the Dirichlet approximation aims to highlight our proposal's flexibility. Figure~\ref{fig:two.examples.a}(d) shows the approximate IM contour based on $K=3$ and counts $X=(8, 10, 7)$.  The exact IM contour is virtually impossible to evaluate, because naive Monte Carlo is slow, the contours are noisy when based on Monte Carlo sample sizes that are too small, and the discrete nature of the data gives it the unusual shape akin to the binomial plot in Figure~\ref{fig:two.examples.a}(a). Here, however, we get a smooth contour approximation in a matter of seconds.  
\end{ex}

Our second goal in this section is to provide a more in-depth illustration of the kind of analysis that can be carried out with the proposed variational-like approximate IM.  We do this in the context of regression modeling for count data.  

\begin{figure}[t]
\begin{center}
\subfigure[Joint contour for $(\Theta_1,\Theta_2)$]{\scalebox{0.53}{\includegraphics{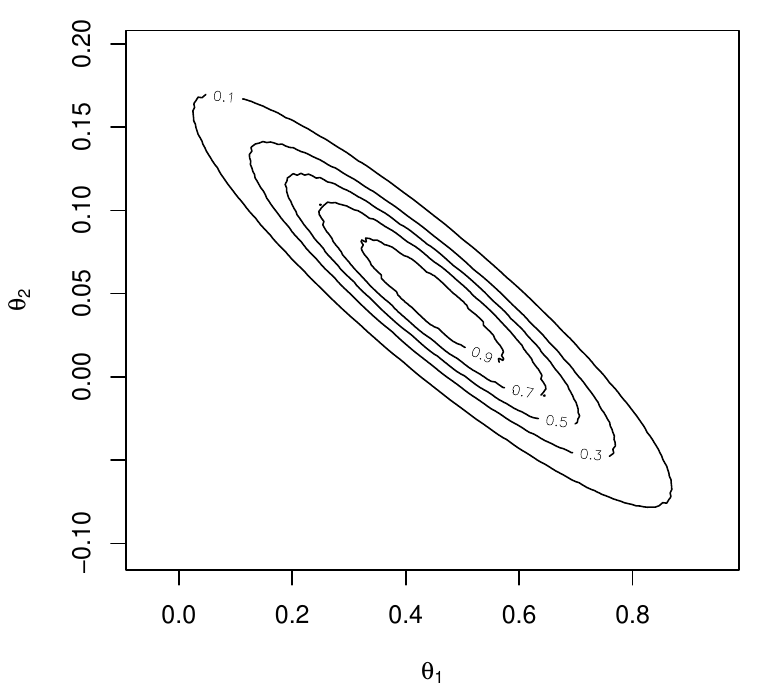}}}
\subfigure[Marginal contour for $\Theta_1$]{\scalebox{0.53}{\includegraphics{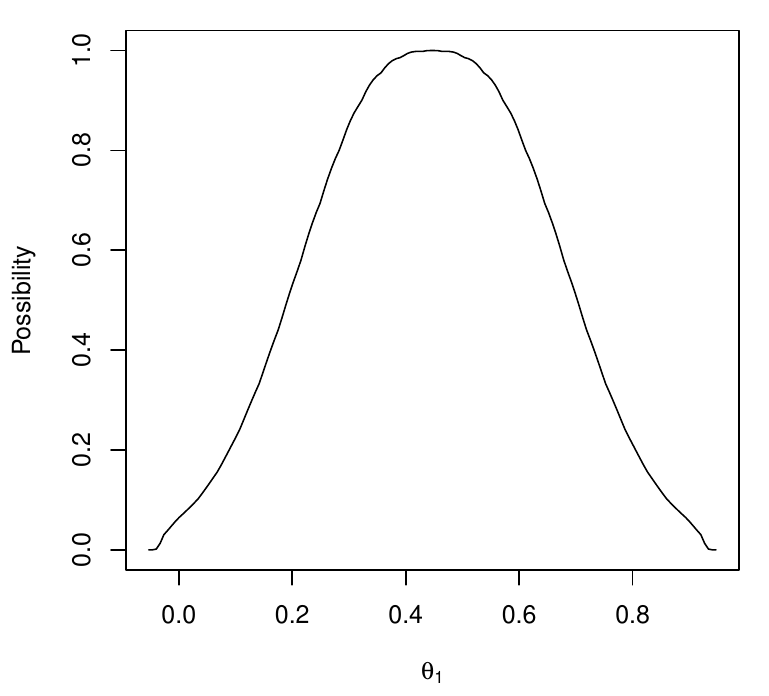}}}
\subfigure[Marginal contour for $\Theta_2$]{\scalebox{0.53}{\includegraphics{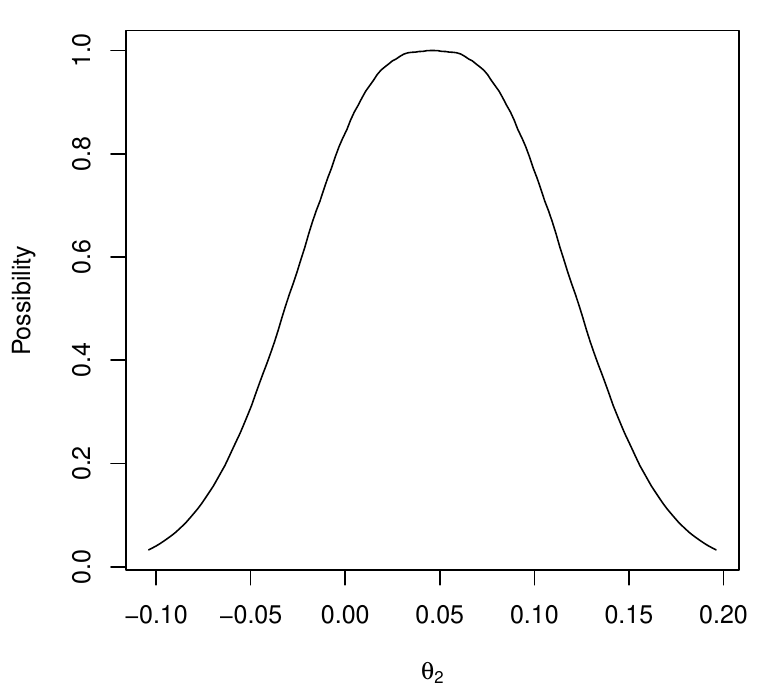}}}
\subfigure[Plot of $\gamma \mapsto \lPi_{X^n}(\Theta_1 > \gamma)$]{\scalebox{0.53}{\includegraphics{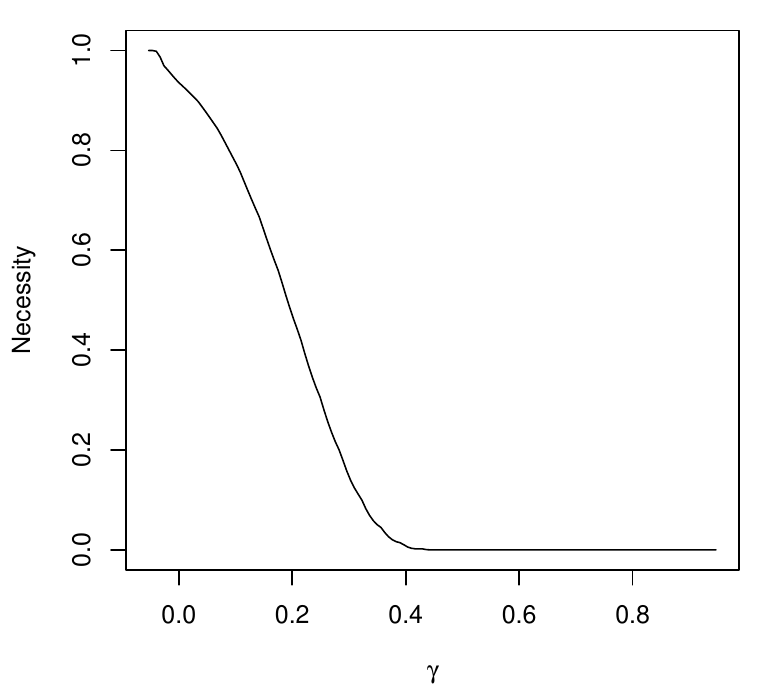}}}
\end{center}
\caption{Results for the Poisson model in Example~\ref{ex:poisson}. Panels~(a), (b) and (c) show the IM's variational-like approximate joint and marginal contours. Panel~(d) shows the necessity measure for hypothesis of the type $\Theta_1 > \gamma$, for $\gamma \geq 0$.}
\label{fig:Poisson}
\end{figure}

\begin{ex}
\label{ex:poisson}
Poisson log-linear models are widely used to analyze the relationship between a count-based discrete response variable and a set of fixed explanatory variables; even if the explanatory variables are not fixed by design, it is almost always assumed that their distribution does not depend on any of the relevant parameters; this makes the explanatory variables ancillary statistics and hence it is customary to condition on their observed values. Let $X_i$ represent the response variable, and $z_{i1}, \ldots, z_{ip}$ denote the $p$ explanatory variables for the $i^{th}$ observation in a sample, $i=1, \ldots,n$.
The Poisson log-linear model assumes a Poisson model for $X_i$, given $z_i=(z_{i1},\ldots,z_{ip})$, independent across $i=1,\ldots,n$, with marginal mass function
\[p_{\lambda_{i}}(x_i) = \lambda_i^{x_i} \, e^{-\lambda_i} \, / \, x_i!, \quad x_i = 0, 1, 2, \ldots, \]
where $\log \lambda_i = \theta_0 + \theta_1 z_{i1} + \theta_2 z_{i2} + \ldots + \theta_p z_{ip}.$
The goal is to quantify uncertainty about the unknown true regression coefficients $\Theta_0, \Theta_1, \ldots, \Theta_p$. 

Consider the data presented in Table~3.2 of \citet{agresti.intro}, obtained from a study of the nesting habits of horseshoe crabs. In this study, each of the $n = 173$ female horseshoe crabs had a male attached to her nest, and the goal was to explore factors influencing whether additional males, referred to as satellites, were present nearby. The response variable $X$ is the number of satellites observed for each female crab. Here, we focus on evaluating two explanatory variables related to the female crab’s size that may influence this response: $z_1$, weight (kg), and $z_2$, shell width (cm).

A variational approximation of the IMs contour was obtained from \eqref{eq:normalappox}, with $\xi$ estimated using $M=200$ Monte Carlo samples, step sizes $w_t = 2(1+t)^{-1}$, and convergence threshold $\eps=0.005$. Perhaps the first question to consider is whether at least one of $z_1$ or $z_2$ has an influence on $X$. To address this, a marginal IM for $(\Theta_1,\Theta_2)$ is constructed, as shown in Figure~\ref{fig:Poisson}(a). Notably, the hypothesis ``$H: \Theta_1 = \Theta_2 = 0$'' has an upper probability close to zero, providing strong evidence that at least one of $\Theta_1$ or $\Theta_2$ differs from zero.  To evaluate the impact of $z_1$ and $z_2$ individually, the respective marginal IMs for $\Theta_1$ and $\Theta_2$ are shown in Figure~\ref{fig:Poisson}(b,c). 
While there is strong evidence supporting ``$\Theta_1>0$,'' the hypothesis that ``$\Theta_2 = 0$'' is highly plausible. Finally, given the apparent evidence for ``$\Theta_1>0$,'' one might ask which hypotheses of the form  ``$H_\gamma : \Theta_1 > \gamma$,'' for $\gamma >0$, are well supported. This question can be addressed using the marginal necessity measures for $H_\gamma$, as shown in Figure~\ref{fig:Poisson}(d). We can see that ``$\Theta_1 > 0.1$'' is well supported, indicating that an additional kilogram increases the average number of satellites per female crab by approximately $10\%$. Importantly, the IM’s uniform validity insures that the probability that even one of the suggestions above is misleading is controllably small.
\end{ex}

\begin{figure}[t]
\begin{center}
\scalebox{0.7}{\includegraphics{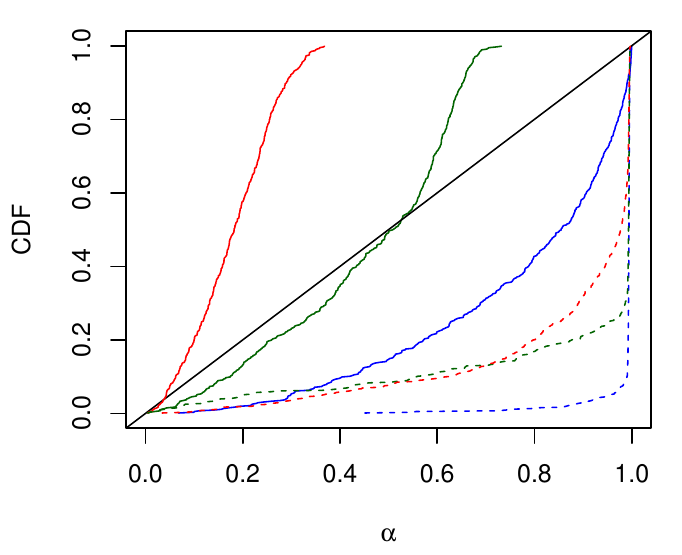}}
\end{center}
\caption{Distribution functions $G_k^\text{\sc im}$ (dashed) and $G_k^\text{\sc bayes}$ (solid) of
the IM possibilities and Bayes posterior probabilities assigned to $H_1:\Theta_1 < 0.5$ (blue), $H_2:0 < \Theta_2 < 0.5$ (green) and $H_3: \Theta_1 > \Theta_2$ (red), based on 500 Monte Carlo samples.}
\label{fig:Bayes}
\end{figure}

To highlight the IM'a calibration guarantees and provide a comparison with an alternative framework, we conduct a simulation study following the setup described above. We generate 500 datasets, each with $n = 25$ observations. The 25 pairs of explanatory variables remain fixed across all datasets and are obtained through a random selection from the 173 pairs in the original dataset described above. Importantly, the selected explanatory variables are scaled so that $\sum_{j=1}^p z_{ij} = 0$ and $p^{-1} \sum_{j=1}^p z_{ij}^2=1$---this scaling does not affect the dependence between the explanatory variables, it ensures that the $\Theta_j$'s are comparable, so that a hypthesis like $H_3$ below is meaningful. The response variable, the $X_i$'s, are independently sampled from $\pois(\Lambda_i)$, with 
\[ \log \Lambda_i = \Theta_0 + \Theta_1 z^*_{i1} + \Theta_2 z^*_{i2},\] 
where $z^*_{i1}$ and $z^*_{i2}$ represent the scaled $z_{i1}$ and $z_{i2}$, respectively,
and $\Theta_0= 1$, $\Theta_1 = 0.25$ and $\Theta_2 = 0.1$. Consider the following three practically relevant hypotheses,
\[H_1:\Theta_1 < 0.5; \quad  \quad H_2:0 < \Theta_2 < 0.5; \quad \quad  H_3: \Theta_1 > \Theta_2 ,\]
concerning different features of $\Theta$. Note that {\em all three hypotheses are true}, i.e., the true $\Theta$ satisfies the constraints specified by these hypotheses. Figure~\ref{fig:Bayes} plots the  distribution function of the possibility assigned to these hypotheses by the IM, obtained via the variational approximation described above:
\[ G_k^\text{\sc im}(\alpha) = \prob_\Theta\{ \uPi_{X^n}(H_k) \leq \alpha \}, \quad \alpha \in (0,1), \quad k=1,2,3. \]
As expected, the IM possibility assignments are well calibrated for all three hypotheses---\eqref{eq:valid.alt} implies that this curve would be below the diagonal line.  Figure~\ref{fig:Bayes} also shows the distribution functions,
\[ G_k^\text{\sc bayes}(\alpha) = \prob_\Theta\{ \prior_{X^n}(H_k) \leq \alpha \}, \quad \alpha \in (0,1), \quad k=1,2,3, \]
of the posterior probabilities $\prior_{X^n}(H_k)$ assigned to the above hypotheses using a Bayesian Poisson regression model, where independent Gaussian priors with mean zero and variance $100^2$ were placed on the regression coefficients. While the posterior probabilities are well calibrated for $H_1$, it is not calibrated for $H_2$ and $H_3$---being above the diagonal line means the posterior probabilities assigned to these true hypotheses are stochastically smaller than $\unif(0,1)$. In particular, $H_3$ gets assigned posterior probabilities less than 0.1 around 20\% of the time. This lack of calibration is undesirable, as it can lead to ``systematically misleading conclusions'' \citep{reid.cox.2014}.

\section{Beyond basic variational-like IMs}
\label{S:beyond}

\subsection{Setup and proposal}

Various tweaks to the basic procedure described in Section~\ref{S:basic} might be considered in order applications to speed up computations.  If there is a computational bottleneck remaining in the proposal from Section~\ref{S:basic}, then it would have to be that the IM's possibility contour must be evaluated at $M$-many points in each iteration of the stochastic approximation.  While this is not impractically expensive in some applications, including those presented above, it could be a substantial burden in other applications.  A related challenge is that a scalar index $\xi$ on the variational family, as we have focused on exclusively so far, may not be sufficiently flexible.  Increasing the flexibility by considering higher-dimensional $\xi$ would likewise increase the computational burden, so care is needed.  Here we aim to address both of the aforementioned challenges.  

The particular modification that we consider here is best suited for cases where the variational family $\credal^\text{var}$ satisfies the following property: for each $\xi \in \Xi$, the $100(1-\alpha)$\% credible set corresponding to $\prior_n^\xi$ can be expressed (or at least neatly summarized) in closed-form.  The idea to be presented here is more general, but, to keep the details as simple and concrete as possible, we will focus our presentation on the case of a Gaussian variational family.  Then the credible sets are ellipsoids in $d$-dimensional space.  

As a generalization of the scalar-$\xi$-indexed Gaussian family introduced in Section~\ref{S:basic}, let $\xi \in \Xi = \RR_{\geq 0}^d$ be a $d$-vector index and take $\credal^\text{var}$ to be the $d$-dimensional Gaussian family with mean vector $\hat\theta_n$, the maximum likelihood estimator, and covariance matrix $J_n(\xi)^{-1}$ defined as follows: take the eigen-decomposition of the observed Fisher information matrix $J_n$ as $J_n = U \Psi U^\top$, and then set 
\[ J_n(\xi) = U \, \text{diag}(\xi^{-1}) \, \Psi \, \text{diag}(\xi^{-1}) \, U^\top. \]
That is, the components of $\xi$ act as multipliers on the singular values of $J_n^{-1}$.  This indeed generalizes our previously-considered Gaussian approximation family since, if $\xi$ is a scalar, then we recover the previous covariance matrix: $J_n(\xi) = J_n / \xi^2$.  Let $C_\alpha^\xi(x^n)$ denote the $100(1-\alpha)$\% credible set corresponding to the distribution $\prior_{n,\alpha}^\xi \in \credal^\text{var}$.  In the present $d$-dimensional Gaussian case, this is an ellipsoid in $\RR^d$, i.e., 
\[ C_\alpha^\xi(x^n) = \{ \theta \in \RR^d: (\theta - \hat\theta_n)^\top J_n(\xi) \, (\theta - \hat\theta_n) \leq \chi_d^2(1-\alpha) \}, \]
where $\chi_d^2(p)$ is the $p^\text{th}$ quantile of the $\chisq(d)$ distribution.  Then we know that $\prior_{n,\alpha}^\xi$ assigns at least probability $1-\alpha$ to the IM's confidence set $C_\alpha(x^n)$ if 
\[ C_\alpha^\xi(x^n) \supseteq C_\alpha(x^n), \]
or, equivalently, if 
\[ \sup_{\theta \not\in C_\alpha^\xi(x^n)} \pi_{n}(\theta) \leq \alpha. \]
For the Gaussian family of approximations we are considering here, the mode of $\pi_n$, the maximum likelihood estimator, is in the interior of each credible set, so the behavior of $\pi_n$ outside $C_\alpha^\xi(x^n)$ is determined by its behavior on the boundary.  Moreover, since equality in the above display implies a near-perfect match between the IM's confidence sets and the posited Gaussian credible sets, a reasonable goal is to find a root to the function 
\[ g_\alpha(\xi) := \max_{\theta \in \partial C_\alpha^\xi(x^n)} \pi_{n}(\theta) - \alpha, \]
where $\partial A$ denotes the boundary of a set $A$.  In designing an iterative algorithm that makes steps towards this root, it is desirable to allow for steps of different magnitudes depending on the direction, and this requires care. For our present Gaussian approximation family, a natural---albeit minimal---representation of the boundary of the credible region $C_\alpha^\xi(x^n)$ is given by the collection of vectors
\begin{equation}
\label{eq:posts}
\vartheta_s^{\xi, \pm} := \hat\theta_n \pm \bigl\{ \chi_d^2(1-\alpha) \, \xi_s \, / \, \psi_s \}^{1/2} \, u_s, \quad s=1,\ldots,d, 
\end{equation}
where $(\psi_s, u_s)$ is the eigenvalue--eigenvector pair corresponding to the $s^\text{th}$ largest eigenvalue.  Then define the vector-valued function $g_\alpha: \Xi \to \RR^d$ with components 
\begin{equation}
\label{eq:ghat.xi}
\hat g_{\alpha,s}(\xi) = \max\{ \pi_n(\vartheta_s^{\xi, +}), \pi_n(\vartheta_s^{\xi,-})\} - \alpha, \quad s=1,\ldots,d.
\end{equation}
At least intuitively, $\hat g_{\alpha,s}(\xi)$ being less than or greater than 0 determines if the credible set $C_\alpha^\xi(x^n)$ is too large or too small, respectively in the $u_s$-direction.  From here, we can apply the same stochastic approximation procedure to determine a sequence $(\xi^{(t)}: t \geq 1)$, this time of $d$-vectors, i.e., 
\[ \xi_s^{(t+1)} = \xi_s^{(t)} + w_{t+1} \, \hat g_{\alpha,s}(\xi^{(t)}), \quad s=1,\ldots,d, \quad t \geq 0, \]
that converges to the root of $\hat g_\alpha$ which we expect to be close to the root of $g_\alpha$.  The details of this new but familiar strategy are presented in Algorithm~\ref{algo:varim2}.  The chief advantage of this new strategy compared to that summarized in Algorithm~\ref{algo:varim} is computational savings: each iteration of stochastic approximation in Algorithm~\ref{algo:varim2} only requires (Monte Carlo) evaluation of the IM contour at $2d$ many points, which would typically be far fewer than then number of Monte Carlo samples $K$ in Algorithm~\ref{algo:varim} needed to get a good approximation of $\prior_{n,\alpha}^\xi\{ \pi_{n}(\Theta) > \alpha\}$.  That is, the computational complexity of this new strategy is $O(Md)$, where $M$ is the Monte Carlo sample size in \eqref{eq:naive} to evaluate $\pi_{x^n}$.


\begin{algorithm}[t]
\SetAlgoLined
requires: data $x^n$, eigen-pairs $(\psi_s, u_s)$, and ability to evaluate $\pi_n=\pi_{x^n}$\; 
initialize: $\alpha$-level, guess $\xi^{(0)}$, step size sequence $(w_t)$, and threshold $\eps > 0$\; 
set: {\tt stop = FALSE}, $t=0$\; 
\While{{\tt !stop}}{
construct the representative points $\{\vartheta_s^{\xi^{(t)}, \pm}: s=1,\ldots,d\}$ as in \eqref{eq:posts}\; 
evaluate $\hat g_{\alpha,s}(\xi_s^{(t)})$ for $s=1,\ldots,d$ as in \eqref{eq:ghat.xi}\;
update $\xi_s^{(t+1)} = \xi_s^{(t)} \pm w_{t+1} \, \hat{g}_{\alpha,s}(\xi_s^{(t)})$ for $s=1,\ldots,d$\;
\eIf{$\max_s |\xi_s^{(t+1)} - \xi_s^{(t)}| < \eps$}{
  $\hat\xi = \xi^{(t+1)}$\;
  {\tt stop = TRUE}\;
}{
  $t \gets t+1$\;
}
}
return $\hat\xi$\; 
\caption{$\alpha$-specific variational approximation of the IM}
\label{algo:varim2}
\end{algorithm}

\subsection{Numerical illustrations}

We will illustrate this new version of the Gaussian variational family and approximation algorithm with three examples.  The first revisits the previous bivariate normal correlation example but with more specific details about timing and accuracy of the approximation; the second is the classical two-parameter gamma model; and the third is a relatively high-dimensional model involving a penalty, which is intended to serve as a gateway into IM solutions to high-dimensional problems.  

\begin{ex}
\label{ex:bivariate.again}
Here we revisit the bivariate normal correlation illustration in Example~\ref{ex:bivariate} above, this time with a quantitative comparison of the computation time and accuracy of the proposed approximation---the version described in Algorithm~\ref{algo:varim2}---and the naive Monte Carlo contour evaluation as in \eqref{eq:naive}.  We follow the setup in Example~\ref{ex:bivariate}, where data sets of varying size $n$ are generated from the standard bivariate normal distribution with true correlation $\Theta=0.5$.  In this case, we generate 100 such data sets of sizes $n=50$, 100, and 200 and, for each data set, we evaluate the IM contour function, say $\hat\pi^\text{naive}$, based on the naive strategy in \eqref{eq:naive} and also based on the newly proposed approximation, say $\hat\pi^\text{approx}$; both are based on $M=500$ Monte Carlo when the contour is evaluated.  In Table~\ref{tab:biv_comp} we compare the relative computation time---defined as the naive strategy's time divided by the proposed strategy's time---to evaluate the contour over a grid of 100 $\theta$ values in $[-1,1]$ and the $L_1$ distance $\int |\hat\pi^\text{approx}(\theta) - \hat\pi^\text{naive}(\theta)| \, d\theta$ between the two; the values are averages over the 100 data sets of each size $n$.  The interpretation here is that the naive strategy is the ``gold-standard'' in the sense that it provides an accurate assessment of the IM contour at each grid point.  So, ideally, the distance of $\hat\pi^\text{approx}$ from $\hat\pi^\text{naive}$ is small but that the former solution is more efficient than the latter's brute-force approach.  As the sample size $n$ increases, however, we know two things:
\begin{itemize}
\item both strategies have computational complexity linear in $n$, but the naive strategy's rate of growth is faster than the proposed approximation's, so the computation time ratio should be roughly constant and greater than 1, and 
\vspace{-2mm}
\item the Gaussian approximation employed in $\hat\pi^\text{approx}$ improves in accuracy, by the results in \citet{imbvm.ext}, so we expect the two contours to merge. 
\end{itemize} 
The results in Table~\ref{tab:biv_comp} confirm these expectations, i.e., the relative computation time holds steady near value 2, so that the naive strategy takes roughly twice as long to compute as the proposed strategy, and the $L_1$ distance between the two contours is decreasing with $n$.  The computation time comparison is favorable for the naive strategy in this case, since it only involves a scalar parameter, but still the performance difference is substantial.
\end{ex}

\begin{table}[t]
    \centering
    \begin{tabular}{ccc}
        \hline
        Sample size, $n$ & Relative time & $L_1$ distance \\
        \hline
        50 & 1.92 & 0.037 \\
        100 & 1.87 & 0.021 \\
        200 & 1.84 & 0.011 \\
        \hline
    \end{tabular}
    \caption{Computation time and accuracy comparisons of the two strategies for evaluating the IM contour in the bivariate normal model: the naive strategy \eqref{eq:naive} and the proposed approximation as described in Algorithm~\ref{algo:varim2}. Based on averages over 100 data sets.}
    \label{tab:biv_comp}
\end{table}

\begin{figure}[t]
\begin{center}
\subfigure[Joint contour for $(\Theta_1, \Theta_2)$]{\scalebox{0.57}{\includegraphics{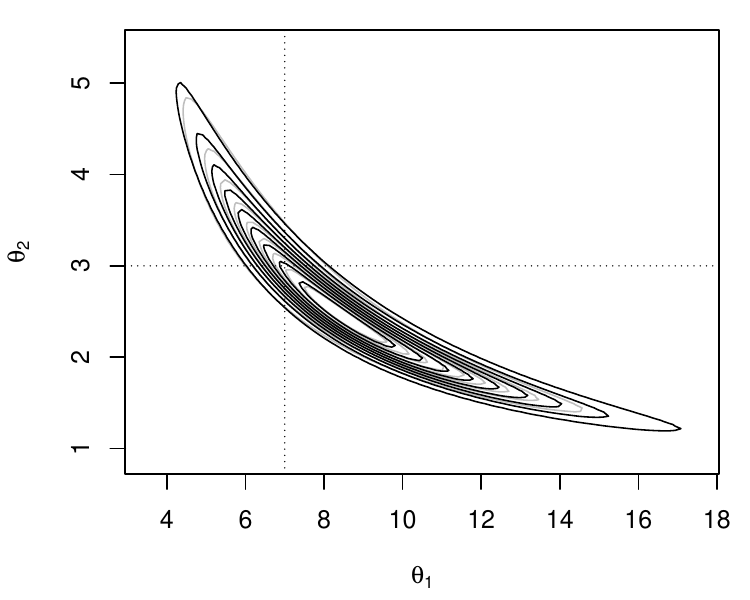}}}
\subfigure[Distribution function of $\pi_{X^n}(\Theta)$]{\scalebox{0.57}{\includegraphics{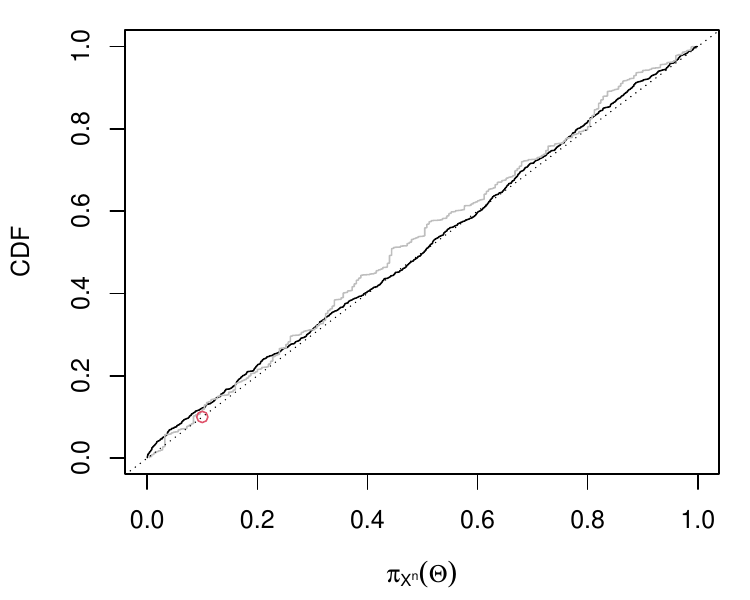}}}
\end{center}
\caption{Results for the gamma model in Example~\ref{ex:gamma}. Panel~(a) shows the approximate IM contour (for log-shape and log-scale) for a single data set; gray in the background shows the relative likelihood contours for comparison. Panel~(b) shows the distribution function for the approximate contour (black) and exact contour (gray) at the true $\Theta$.}
\label{fig:gamma}
\end{figure}

\begin{ex}
\label{ex:gamma}
Suppose $X^n$ consists of an iid sample of size $n$ from a gamma distribution with shape parameter $\Theta_1$ and scale parameter $\Theta_2$.  We simulated data of size $n=25$, with $\Theta_1=7$ and $\Theta_2=3$, and we plot the approximate IM contour for $(\Theta_1, \Theta_2)$ in Figure~\ref{fig:gamma}(a).  This contour is constructed by first building a Gaussian approximation of the contour for $(\log\Theta_1, \log\Theta_2)$ and then mapping the approximation back to the $(\Theta_1, \Theta_2)$ space.  The parameters' non-negativity constraints are gone when mapped to the log-scale, which improves the quality of the Gaussian approximation; the approximation is worse when applied directly on the $(\theta_1,\theta_2)$ space.  For comparison, the contour of the relative likelihood is also shown, and its similarity to that of the Gaussian contour implies that the latter is a good approximation to the exact IM contour, which is rather expensive to compute.   Indeed, Example~1 in \citet{gim} considers exactly the same simulation setting and he also obtains a banana-shaped contour similar to the one in Figure~\ref{fig:gamma}(a). 

We also repeated the above example 1000 times and Figure~\ref{fig:gamma}(b) shows the distribution function of the random variable $\pi_{X^n}(\Theta)$, based on both the exact contour and the Gaussian approximation.  That with the exact contour is the $\unif(0,1)$ distribution function (modulo Monte Carlo sampling variation) and that based on the Gaussian approximation is empirically indistinguishable from $\unif(0,1)$ across the full range.  
\end{ex}

\begin{figure}[t]
\begin{center}
\subfigure[Plot of the estimated $\xi$ values]{\scalebox{0.57}{\includegraphics{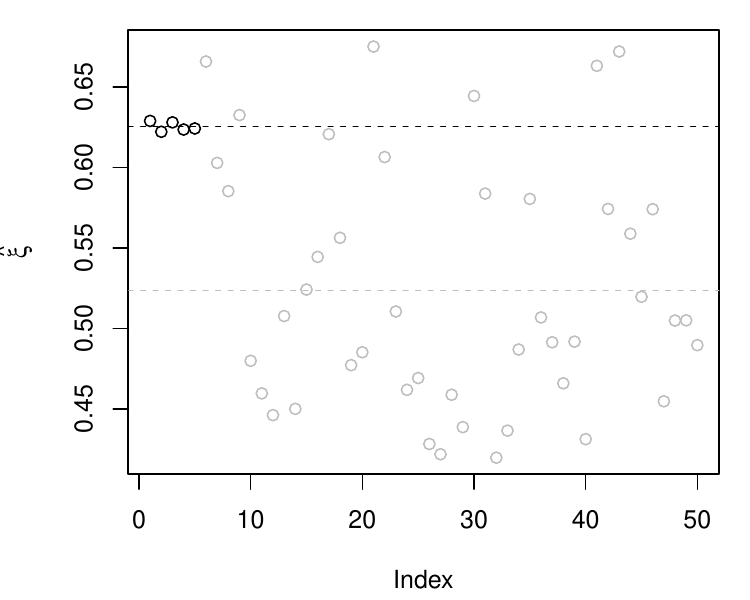}}}
\subfigure[Distribution function of $\pi_{X^n}(\Theta)$]{\scalebox{0.57}{\includegraphics{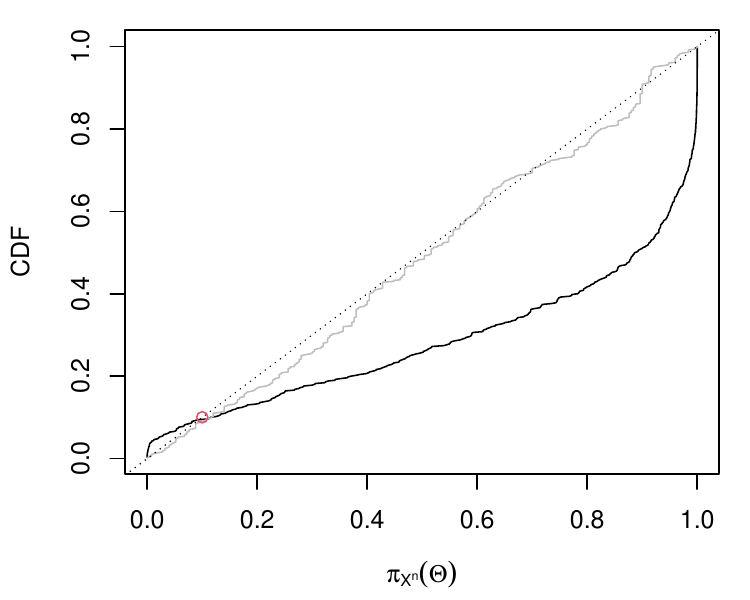}}}
\end{center}
\caption{Results from the many-normal-means model in Example~\ref{ex:lasso}. Panel~(a) plots the estimated $\xi$ values based on a single data set; black and gray horizontal lines correspond to the averages of the black/signal and gray/noise points. Panel~(b) shows the distribution function for the approximate contour (black) and exact contour (gray) at the true $\Theta$.}
\label{fig:lasso}
\end{figure}

\begin{ex}
\label{ex:lasso}
This simplest and most canonical high-dimensional inference example is the {\em many-normal-means problem}, which goes back to classical papers such as \citet{stein1956}, \citet{james.stein.1961}, \citet{brown1971}, and others.  The model assumes that the data $X^n$ consists of $n$ independent---but not iid---observations, with $X_i \sim \nm(\Theta_i, \sigma^2)$, where $\sigma^2$ is assumed to be known but the vector $\Theta=(\Theta_1,\ldots,\Theta_n)$ is unknown and to be inferred.  Roughly, the message in the references above is that the best unbiased estimator $\hat\theta = X^n$, also the maximum likelihood estimator, is inadmissible (under squared error loss).  This result sparked research efforts on penalized estimation, including the now-famous lasso \citep[e.g.,][]{tibshirani1996, tibshirani2011}.  Following along these lines, and in the spirit of the examples in Section~\ref{S:extensions} above, we propose here to work with a relative penalized likelihood 
\[ R_\lambda^\text{\sc pen}(x^n,\theta) = \frac{L_{x^n}^{\text{\sc pen},\lambda}(\theta)}{\max_\vartheta L_{x^n}^{\text{\sc pen},\lambda}(\vartheta)}, \quad \theta \in \TT = \RR^n, \]
where 
\[ L_{x^n}^{\text{\sc pen},\lambda}(\theta) = L_{x^n}(\theta) \, e^{-\lambda \|\theta\|_1} = \exp\bigl\{ -\tfrac{1}{2\sigma^2} \|x^n - \theta\|_2^2 - \lambda \|\theta\|_1 \bigr\}, \]
is the penalized likelihood function, with $\|\cdot\|_1$ and $\|\cdot\|_2$ the $\ell_1$- and $\ell_2$-norms, respectively, and $\lambda > 0$ a tuning parameter to be discussed below.  For this relatively simple high-dimensional model, the solution to the optimization problem in the denominator above, which is the lasso estimator, can be solved in closed-form: if the vector 
\[ \hat\theta = \arg\max_\vartheta L_{x^n}^{\text{\sc pen},\lambda}(\vartheta) \]
is the solution, then its entries are given by 
\[ \hat\theta_i = \text{sign}(x_i) \, \max( |x_i| - \lambda, 0 ), \quad i=1,\ldots,n. \]
That is, the lasso penalization shrinks the maximum likelihood estimator $X_i$ of $\Theta_i$ closer to 0 and, in fact, if the magnitude of $X_i$ is less than the specified $\lambda$, then the estimator is exactly 0.  Of course, the choice of $\lambda$ is critical, and here we will take it to be $\lambda = (\sigma^2 \log n)^{1/2}$.  In any case, we can construct a possibilistic IM contour as 
\[ \pi_{x^n}^{\text{\sc pen},\lambda}(\theta) = \prob_\theta\{ R_\lambda^\text{\sc pen}(X^n,\theta) \leq R_\lambda^\text{\sc pen}(x^n,\theta)\}, \quad \theta \in \TT. \]
Even though the relative penalized likelihood is available in closed-form, Monte Carlo methods are needed to evaluate this contour.  And when the dimension $n$ is even moderately large, carrying out these computations on a grid in $n$-dimensional space that is sufficiently fine to obtain, say, confidence sets for $\Theta$ is practically impossible.  But the variational approximation described above offers a computationally more efficient alternative to naive Monte Carlo that can handle moderate to large $n$.

Here we follow the strategy suggested above, i.e., with a Gaussian approximation having mean $\hat\theta$ equal to the lasso or maximum penalized likelihood estimator and $n \times n$ covariance matrix $J_n(\xi)^{-1}$ indexed by the $n$-vector $\xi$; in this case, the initial $J_n$ is taken to be the no-penalty information matrix $J_n = \sigma^{-2} I_n$, which is proportional to the identity matrix.  The intuition here is that the coordinate-specific adjustment factors, $\xi_i$, will allow the Gaussian approximation to adapt in some way to sparsity in the true signal $\Theta$.  For our illustration, we consider $n=50$ and a true $\Theta$ whose first five entries are equal to 5 and last 45 entries equal to 0, i.e., only 10\% of the coordinates of $X$ contain a signal, the remaining 90\% is just noise.  We also fixed $\alpha=0.1$ for the approximation.  For a single data set, Figure~\ref{fig:lasso}(a) displays the estimates $\hat\xi$ obtained at convergence of the proposed stochastic approximation updates.  The points in black correspond to signals (non-zero true means) and the points in gray correspond to noise.  The key observation is that $\xi$ values corresponding to signal tend to be larger than those corresponding to noise; there is virtually no variability in the signal cases, but substantial variability in the noise cases.  That the $\xi$ values tend to be smaller in the noise case is expected, since much less spread in the IM's possibility contour is needed around those means that are clearly 0. We repeated the above simulation 1000 times and plotted the distribution function of the exact (using naive Monte Carlo) and approximate (using the Gaussian variational family) IM contour at the true $\Theta$.  Again, the exact contour at $\Theta$ is $\unif(0,1)$ distributed, and the results in Figure~\ref{fig:lasso}(b).  The Gaussian approximation is only designed to be calibrated at level $\alpha=0.1$, which is clearly achieves; it is a bit too agressive at lower levels and conservative at the higher levels.  Further investigation into this proposal in high-dimensional problems will be reported elsewhere. 
\end{ex}

\section{Nuisance parameter problems}
\label{S:extensions}

\subsection{Parametric}
\label{SS:parametric}

The perspective taken above was that there is an unknown model parameter $\Theta$ and the primary goal is to quantify uncertainty about $\Theta$ in its entirety given the observed data $X^n=x^n$ from the model.  Of course, quantification of uncertainty about $\Theta$ implies quantification of uncertainty about any feature $\Phi = g(\Theta)$ as described in Section~\ref{S:background} above.  If, however, the sole objective is quantifying uncertainty about a particular $\Phi = g(\Theta)$, then it is natural to ask if we can do better than first quantifying uncertainty about the full $\Theta$ and then deriving the corresponding results for $\Phi$.  There are opportunities for efficiency gain, but this requires eliminating the nuisance parameters, i.e., those aspects of $\Theta$ that are, in some sense, complementary or orthogonal to $\Phi$.  One fairly general strategy for eliminating nuisance parameters is {\em profiling} \citep[e.g.,][]{martin.partial3, SeveriniWong:1992, sprott2000}, described below.  

It is perhaps no surprise that, while use of the relative likelihood in the IM construction presented in Section~\ref{S:background} is very natural and, in some sense, ``optimal,'' it is not the only option.  For cases involving nuisance parameters, one strategy is to replace the relative likelihood in \eqref{eq:contour} with a surrogate, namely, the relative profile likelihood
\begin{equation}
\label{eq:pi.profile}
R^\text{\sc pr}(x^n, \phi) = \frac{\sup_{\theta: g(\theta) = \phi} L_{x^n}(\theta)}{\sup_{\vartheta \in \TT} L_{x^n}(\vartheta)}. 
\end{equation}
Note that this is not based on a genuine likelihood function, i.e., there is no statistical model for $X^n$, depending solely on the unknown parameter $\Phi$, for which $\phi \mapsto \sup_{\theta: g(\theta)=\phi} L_{x^n}(\theta)$ is the corresponding likelihood function.  Despite not being a genuine likelihood, the profile likelihood typically behaves very much like the usual likelihood asymptotically and, therefore, ``may be used to a considerable extent as a full likelihood'' \citep[][p.~449]{murphy.vaart.2000}.  In our present case, if we proceed with the relative profile likelihood in place of the relative likelihood in \eqref{eq:contour}, then the marginal possibilistic IM has contour function 
\[ \pi_{x^n}^\text{\sc pr}(\phi) = \sup_{\theta: g(\theta)=\phi} \prob_\theta\bigl\{ R^\text{\sc pr}(X^n, \phi) \leq R^\text{\sc pr}(x^n, \phi) \bigr\}, \quad \phi \in g(\TT). \]
The advantage of this construction is that it tends to be more efficient than the naive elimination of nuisance parameters presented in Section~\ref{S:background}; see, e.g., \citet{imbvm.ext}.  The outer supremum above is the result of $\Phi$ not being the full parameter of the posited model; see \citet[][Sec.~3.2]{martin.partial3} for more details.  It is often the case that the probability on the right-hand side above is approximately constant in $\theta$ with $g(\theta)=\phi$, but one cannot count on this---the supremum needs to be evaluated, unfortunately, in order to ensure the IM's strong validity properties hold marginally.  

For concreteness, we will focus on a deceptively-challenging problem, namely, efficient inference on the mean of a gamma distribution.  Roughly, the gamma mean is a highly non-linear function of the shape and scale parameters, which makes the classical, first-order asymptotic approximations of the sampling distribution of the maximum likelihood estimator rather poor in finite samples.  For this reason, the gamma mean problem has received considerable attention, with focus on deriving asymptotic approximations with higher-order accuracy; we refer the reader to \citet{fraser.reid.wong.1997} for further details.  \citet{immarg} presented an exact IM solution to the gamma mean problem and, more recently, a profile-based possibilistic IM solution was presented in \citet[][Example~6]{martin.partial3} and shown to be superior to various existing methods.  Here our focus is demonstrating the quality of the variational approximation in this new context.  

\begin{figure}[t]
\begin{center}
\scalebox{0.7}{\includegraphics{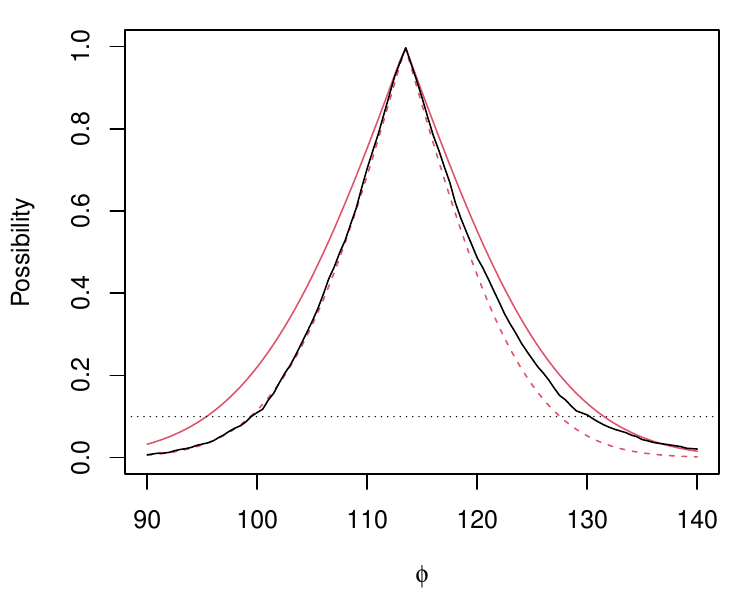}}
\end{center}
\caption{Possibility contour plots for the gamma model in Example~\ref{ex:gamma.mean}.  Black line shows the exact profile-based marginal IM contour, dashed red is the naive Gaussian approximation ($\xi = 1$), and solid red is the Gaussian approximation with the strategically-chosen scale factor $\hat\xi=1.28$ based on $\alpha=0.1$.}
\label{fig:gamma.mean}
\end{figure}

\begin{ex}
\label{ex:gamma.mean}
Let the gamma model be indexed by $\theta = (\theta_1, \theta_2)$, where $\theta_1$ and $\theta_2$ represent the (positive) shape and scale parameters, respectively.  There are no closed-form expressions for the maximum likelihood estimators, $\hat\theta_1$ and $\hat\theta_2$, in the gamma model, but one can readily maximize the likelihood numerically to find them; one can also obtain the observed Fisher information matrix $J$, numerically or analytically.  For the profile likelihood, it may help to reparametrize the model in terms of the mean parameter $\Phi = \Theta_1\Theta_2$ and the shape parameter $\Theta_1$.  Write the density in this new parametrization as 
\[ p_{\theta_1, \phi}(x) = \frac{1}{\Gamma(\theta_1) \, (\phi / \theta_1)^{\theta_1}} \, x^{\theta_1 - 1} \, e^{-\theta_1 x/\phi}, \quad x > 0. \]
In this form, the likelihood based on data $X^n$ could be maximized numerically over $\theta_1$ for any fixed $\phi$, thus yielding the (relative) profile likelihood.  

\citet{fraser.reid.wong.1997} presented an example where $n=20$ mice were exposed to 240 rads of gamma radiation and their survival times recorded.  A plot of the exact profiling-based marginal possibilistic IM contour is shown (black line) in Figure~\ref{fig:gamma.mean}.  This is relatively expensive to compute since, at each $\phi$ point on the grid, our Monte Carlo approximation needs to be optimized over different $\theta_1$ values.  For comparison, we consider a Gaussian possibility contour with mean $\hat\phi = \hat\theta_1 \hat\theta_2$ and variance $\xi^2 \, \dot g(\hat\theta)^\top J^{-1} \dot g(\hat\theta)$, where $g(\theta) = \theta_1\theta_2$ and the gradient is $\dot g(\theta) = (\theta_2, \theta_1)^\top$.  Figure~\ref{fig:gamma.mean} shows the Gaussian approximation with $\xi=1$ as discussed in \citet{imbvm.ext} and with $\hat\xi=1.28$ as identified based on our variational approximation from Section~\ref{S:beyond}.  This approximation takes only a fraction of a second to obtain, and we find that, as intended, it closely matches the exact contour at the target level $\alpha=0.1$ on the right (heavy) side and is a bit conservative on the left (thin) side.  Clearly the basic large-sample Gaussian approximation is too narrow in the right tail, confirming the claim above that the first-order asymptotic theory provides relatively poor approximations for smaller sample sizes.  Our variational approximation, on the other hand, appropriately adjusts to match the exact contour in certain places while being a bit cautious or conservative in others.  
\end{ex}

\subsection{Nonparametric}

A nonparametric problem is one in which the underlying distribution $\prob$ is not assumed to have a particular form indexed by a finite-dimensional parameter.  In some applications, it is $\prob$ itself (or, e.g., its density function) that is the quantity of interest and, in other case, there is some (finite-dimensional) feature or functional $\Theta$ of $\prob$ that is to be inferred.  Our focus here is on the latter case, so it also fits the general mold of a problem involving nuisance parameters, since all of what is left of $\prob$ once $\Theta$ has been accounted for would be considered ``nuisance'' and to-be-eliminated.  

At least in principle, one can approach this problem in a manner similar to that in the parametric case above, i.e., via profiling out those nuisance parts of $\prob$.  Recall that the goal of profiling is to reduce the dimension, so that the compatibility of data with candidate values of the quantity of interest could be directly assessed.  Since it is typically the case that the quantity $\Theta$ of interest has some real-world interpretation, there is an opportunity to leverage that interpretation for the aforementioned compatibility assessment without the need of profiling.  This is the approach taken in \citet{cella.martin.imrisk}, building on the classical work in M-estimation \citep[e.g.,][]{huber1981, boos.stefansky.m-estimation} and less-classical work on Gibbs posteriors \citep[e.g.,][]{martin.syring.chapter2022, grunwald.mehta.rates, bissiri.holmes.walker.2016, zhang2006b}, which we summarize briefly below.  


Let data $X^n = (X_1, \ldots, X_n)$ consist of independent and identically distributed components with $X_i \sim \prob$, where nothing is known or assumed about $\prob$.  In this more general case, the unknown quantity of interest $\Theta=\Theta(\prob)$ is a functional of the underlying distribution. Examples include quantiles and moments of $\prob$. Suppose $\Theta$ can be expressed as the minimizer of a {\em risk} or {\em expected loss} function. That is, assume that there exists a loss function $(x,\theta) \mapsto \text{\sc loss}_\theta(x)$ such that 
\[ \Theta = \arg\min_{\theta \in \TT} \rho(\theta), \]
where $\rho(\theta) = \int \text{\sc loss}_\theta(x) \, \prob(dx)$ is the risk or expected loss function.  An empirical version of the risk function replaces expectation with respect to $\prob$ by averaging over observations: 
\[ \hat\rho_{x^n}(\theta) = \frac1n \sum_{i=1}^n \text{\sc loss}_\theta(x_i), \quad \theta \in \TT. \]
Then the empirical risk minimizer, $\hat\theta_{x^n} = \arg\min_\theta \hat\rho_{x^n}(\theta)$, is the natural estimator of the risk minimizer $\Theta$.  From here, we can mimic the relative likelihood construction with an empirical-risk version 
\[ R^\text{\sc er}(x^n, \theta) = \exp\bigl[ -\{ \hat\rho_{x^n}(\theta) - \hat\rho_{x^n}(\hat\theta_{x^n}) \} \bigr], \quad \theta \in \TT. \]
Clearly, values of $\theta$ such that $R^\text{\sc er}(x^n,\theta)$ is close to 1 are more compatible with data $x^n$ than values of $\theta$ such that it is close to 0.  Note that no profiling over the infinite-dimensional nuisance parameters is required in evaluating $R^\text{\sc er}$.  

The same validification step can be taken to construct a possibilistic IM:
\begin{equation}
\label{eq:nonpIM}
\pi^\text{\sc er}_{x^n}(\theta) = \sup_{\prob: \prob \rightsquigarrow \theta} \prob \bigl\{ R^\text{\sc er}(X^n,\theta) \leq R^\text{\sc er}(x^n,\theta) \bigr\}, \quad \theta \in \TT.  
\end{equation}
The outer supremum, analogous to that in \eqref{eq:pi.profile}, maximizes over all those probability distributions $\prob$ such that the relevant feature $\Theta$ of $\prob$ takes value $\theta$.  This supremum appears because the distribution of $R^\text{\sc er}(X^n,\theta)$ obviously depends on the underlying $\prob$, but this is unknown.  This puts direct evaluation of the IM contour based on naive Monte Carlo out of reach.  Fortunately, validity only requires that certain calibration holds for the one true $\prob$, which provides a shortcut.  \citet{cella.martin.imrisk} proposed to replace ``iid sampling from $\prob$ for all $\theta$-compatible $\prob$'' with iid sampling from the empirical distribution, which is a good estimator of the one true $\prob$.  This amounts to using the {\em bootstrap} \citep[e.g.,][]{efron1979, efrontibshirani1993, davison.hinkley.1997} to approximate the above contour, and Cella and Martin prove that the corresponding IM is asymptotically valid.  Here, we will demonstrate that the proposed variational-like IMs can provide a good approximation for this bootstrap-based contour.

\begin{figure}[t]
\begin{center}
\subfigure[Plausibility contours]{\scalebox{0.53}{\includegraphics{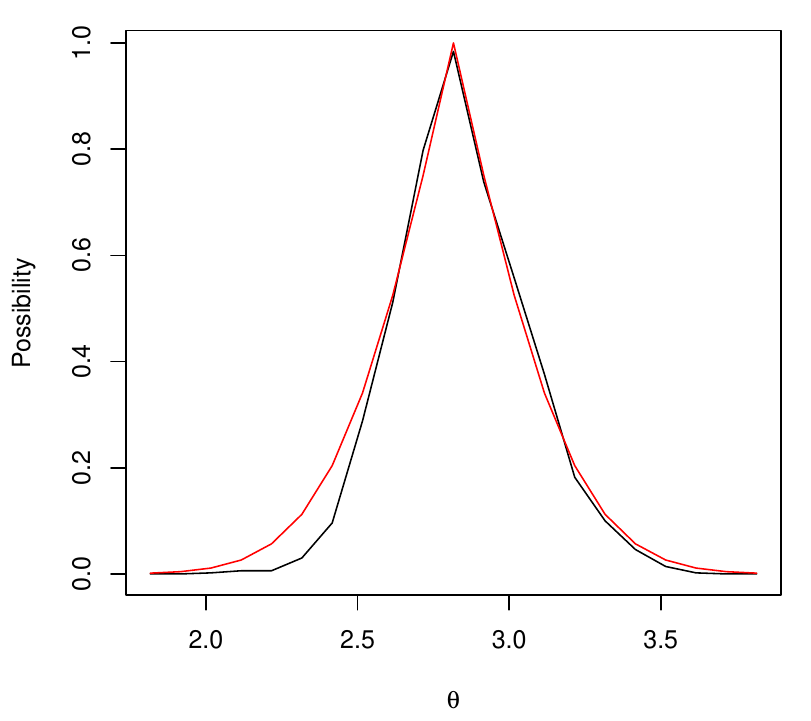}}}
\subfigure[Distribution function of $\pi_{X^n}(\Theta)$]{\scalebox{0.53}{\includegraphics{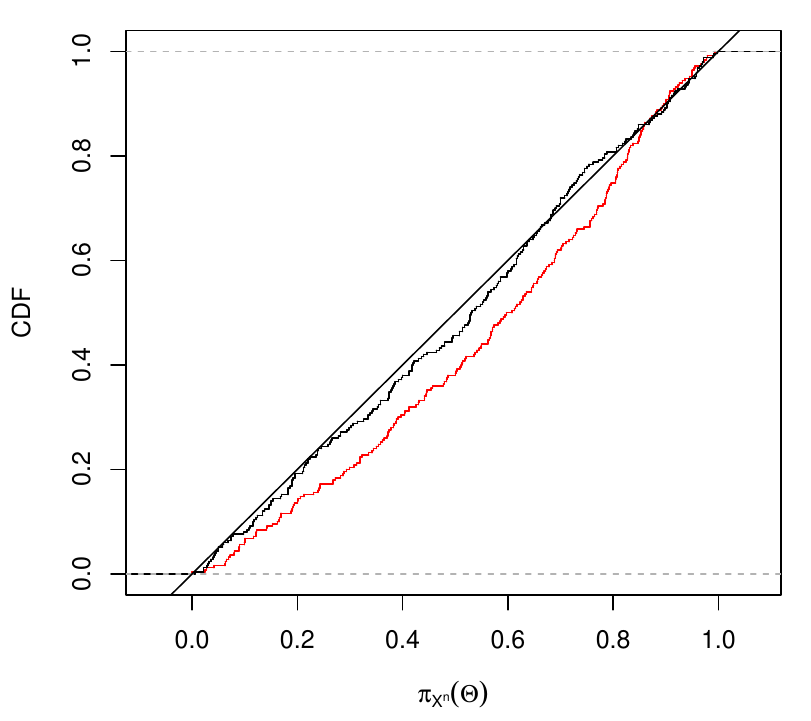}}}
\end{center}
\caption{Results from Example~\ref{ex:quantile} regarding inference on the first quartile. Panel~(a) shows the bootstrap-based contour (black) and its variational approximation (red). Panel~(b) shows the distribution function of both contours at the true $\Theta$.}
\label{fig:Quantile}
\end{figure}

\begin{ex}
\label{ex:quantile}
Suppose interest in the $\tau$-th quantile of a distribution $\prob$, i.e., the exact point $\Theta=\Theta^{(\tau)}$ such that $\prob(X \leq \Theta^{(\tau)}) = \tau$, for $\tau \in (0,1)$. 
The key component in the nonparametric IM construction above is selecting an appropriate loss function. For quantile estimation, it is well-known that the loss function is given by
\[ \text{\sc loss}_\theta(x) = \tfrac12 \bigl\{ (|x-\theta| - x) + (1-2\tau)\theta \bigr\}.  \]
Figure~\ref{fig:Quantile}(a) shows the bootstrap approximation of the IM contour in \eqref{eq:nonpIM}, computed using 500 bootstrap samples, for a dataset of size 
$n=100$ and $\tau=0.25$. We chose to work with the normal family for the variational approximation, due to the well known asymptotic normality of the empirical risk minimizer (sample quantile) $\hat \theta_n^{(\tau)}$:
\[ n^{1/2}(\hat \theta_n^{(\tau)} - \Theta^{(\tau)}) \to \nm\left(0,\frac{\tau(1-\tau)}{p(\Theta^{(\tau)})^2}\right), \quad \text{in distribution}, \]
where $p$ represents the common density function associated with the underlying true distribution $\prob$.  More specifically, our chosen Gaussian variational family $\credal^\text{var}$ has mean $\hat\theta_n^{(0.25)}$ and variance 
\[ \frac{0.25 \times 0.75}{n \, \xi^2 \, \hat p(\hat \theta_n^{(0.25)})^2}, \]
where $\hat p$ denotes a kernel density estimate of $p$ based on the observed data. The same settings as in Section~\ref{S:illustrations} were applied, with $\xi$ estimated using $M=200$ Monte Carlo samples, step sizes $w_t = 2(1+t)^{-1}$, $\alpha=0.1$  and convergence threshold $\eps = 0.005.$ Note how the variational approximation is perfect except for small levels of $\alpha$ on the left side, where it is a bit conservative. 

To verify that the variational approach provides approximate validity in nonparametric settings, a simulation study was performed by repeating the above scenario 250 times. For each dataset, the approximated contour was evaluated at $\Theta = 2.53$, which corresponds approximately to the first quartile when $\prob$ follows a $\gam(4,1)$ distribution. The empirical distribution of this contour is shown in Figure~\ref{fig:Quantile}(b), demonstrating that approximate validity is indeed achieved.
\end{ex}

\subsection{Semiparametric}

A middle-ground between the parametric and nonparametric case described in the previous two subsections is what are called {\em semiparametric} problems, i.e., those with parametric and nonparametric parts.  Perhaps the simplest example is a linear regression model with an unspecified error distribution: the linear mean function is the parametric part and the error distribution is the nonparametric part.  Below we will focus on a censored-data with a semiparametric model but, of course, other examples are possible; see, e.g., \citet{bickel1998}, \citet{tsiatis2006}, and \citet{kosorok.book} for more details.  

The same profiling strategy described in Section~\ref{SS:parametric} above can be carried out for semiparametric models too; \citet{murphy.vaart.2000} is an important reference.  For concreteness, we will consider a common situation involving censored data.  That is, suppose we are measuring, say, the concentration of a particular  chemical in the soil, but our measurement instrument has a lower detectability limit, i.e., concentrations below this limit are undetectable.  In such cases, the concentrations are (left) censored.  We might have a parametric model in mind for the measured concentrations, but the censoring corrupts the data and ultimately changes that model.  Let $Y_i$ denote the actual chemical concentration at site $i$, which we may or may not observe; the $Y_i$'s are assigned a statistical model $\{\prob_\theta: \theta \in \TT\}$, and the true-but-unknown value $\Theta$ of the posited model parameter is to be inferred.  Let $C_i$ denote the censoring level, which we will assume---without loss of generality---to be subject to sampling variability, i.e., the $C_i$'s are random variables.  Then the observed data $X^n$ consists of iid pairs $X_i = (Z_i, T_i)$, where 
\begin{equation}
\label{eq:censoring.rule}
Z_i = \max( Y_i, C_i ) \quad \text{and} \quad T_i = 1(Y_i \geq C_i), \quad i=1,\ldots,n. 
\end{equation}
The goal is to infer the unknown $\Theta$ in the model for the concentrations, but now the measurements are corrupted by censoring.  The augmented model for the corrupted data has likelihood function given by 
\[ L_{x^n}(\theta, G) = \prod_{i=1}^n g(z_i)^{1-t_i} \, G(z_i)^{t_i} \times \prod_{i=1}^n p_\theta(z_i)^{t_i} \, P_\theta(z_i)^{1-t_i}, \]
depending on both the generic value $\theta$ of the true unknown model parameter $\Theta$ for the concentrations and on the generic value $G$ of the true unknown censoring level distribution $\mathsf{G}$.  In the above expression, $g$ and $p_\theta$ are density functions for the censoring and concentration distributions, and $G$ and $P_\theta$ are the corresponding cumulative distribution functions.  Now it should be clear why this is a semiparametric model: in addition to the obvious parametric model, there is the nonparametric model for the censoring levels.  

A distinguishable feature of this semiparametric model is that the likelihood function is ``separable,'' i.e., it is a product of terms involving $\theta$ and terms involving $G$.  Consequently, if we were to optimize over $G$ and then form the relative profile likelihood ratio, then the part involving the optimization over $G$ cancels out.  This implies that we can simply ignore the part with $G$ and work with the relative profile likelihood of the form 
\[ R^\text{\sc pr}(x^n, \theta) = \frac{\prod_{i=1}^n p_\theta(z_i)^{t_i} \, P_\theta(z_i)^{1-t_i}}{\prod_{i=1}^n p_{\hat\theta}(z_i)^{t_i} \, P_{\hat\theta}(z_i)^{1-t_i}}, \quad \theta \in \TT, \]
where $\hat\theta$ is the maximum likelihood estimator of $\Theta$.  The {\em distribution} of the relative profile likelihood still depends on the nuisance parameter $\mathsf{G}$, so when we define the possibilistic IM contour by validifying the relative profile likelihood, we get 
\[ \pi_{x^n}^\text{\sc pr}(\theta) = \sup_G \prob_{\theta, G} \bigl\{ R^\text{\sc pr}(X^n, \theta) \leq R^\text{\sc pr}(x^n, \theta) \bigr\}, \quad \theta \in \TT. \]
Similar to the strategy described above (from \citealt{cella.martin.imrisk}) for the nonparametric problem, \citet{imcens} proposed a novel strategy wherein a variation on the Kaplan--Meier estimator \citep[e.g.,][]{kaplanmeier, km.book} is used to obtain a $\widehat G$, and then the contour is approximated by 
\begin{equation}
\label{eq:censored.boot.pl}
\hat\pi_{x^n}^\text{\sc pr}(\theta) = \prob_{\theta, \widehat G} \bigl\{ R^\text{\sc pr}(X^n, \theta) \leq R^\text{\sc pr}(x^n, \theta) \bigr\}, \quad \theta \in \TT. 
\end{equation}
Evaluation of the right-hand side via Monte Carlo boils down to sampling censoring levels from $\widehat G$, sampling concentration levels from $\prob_\theta$, and then constructing new data sets according to \eqref{eq:censoring.rule}.  While this procedure is conceptually relatively simple, naive implementation over a sufficiently fine grid of $\theta$ values is rather expensive.  Fortunately, our proposed variational-like approximation from Section~\ref{S:beyond} can be readily applied, quickly producing an approximate contour in closed-form.  

\begin{figure}[t]
\begin{center}
\subfigure[Estimated distribution function, $\widehat G$]{\scalebox{0.57}{\includegraphics{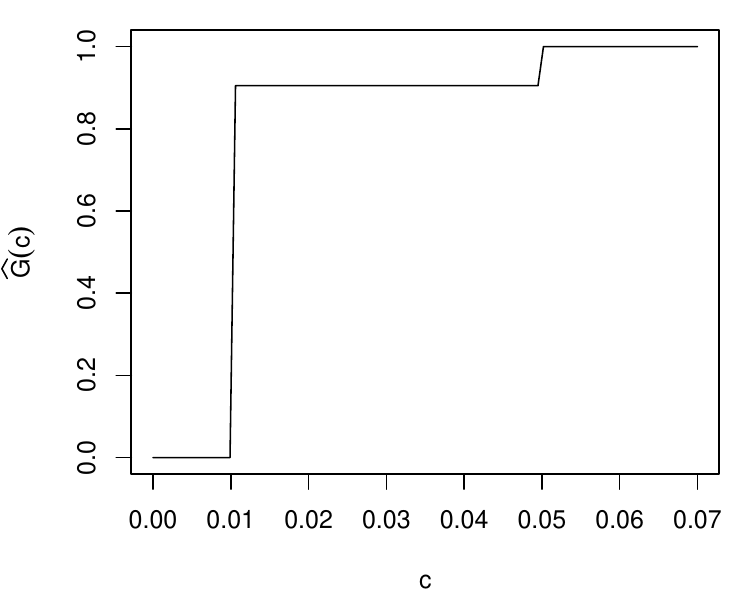}}}
\subfigure[Possibility contour]{\scalebox{0.57}{\includegraphics{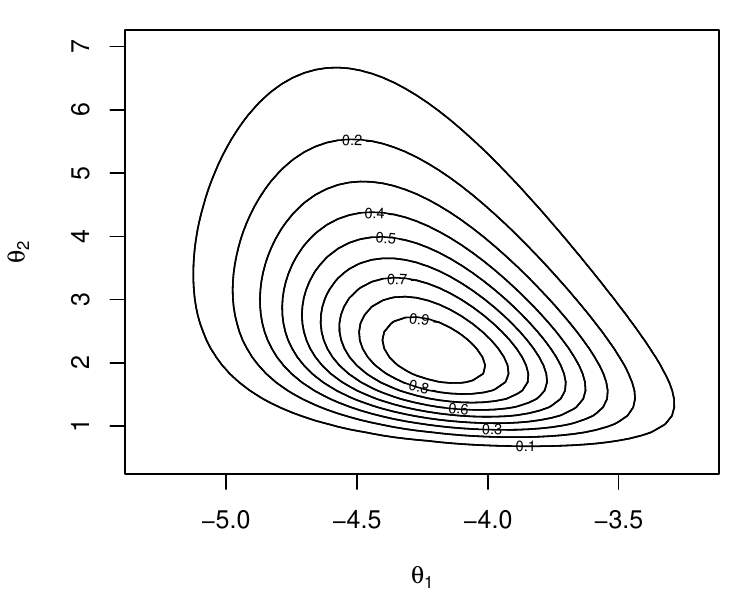}}}
\end{center}
\caption{Results for left-censored data analysis in Example~\ref{ex:censored}.  Panel~(a) shows the Kaplan--Meier estimator of the censoring distribution, which happens to be supported on only two points \citep[see][Sec.~4.3]{imcens}. Panel~(b) shows the IM's variational-like approximate possibility contour.}
\label{fig:censored}
\end{figure}

\begin{ex}
\label{ex:censored}
For illustration, we use data on Atrazine concentration collected from a well in Nebraska.  The data consists of $n=24$ observations subjected to random left-censoring as described above.  This is a rather extreme case where nearly half (11) of the 24 observations are censored, but previous studies indicate that a log-normal model for the Atrazine concentrations is appropriate \citep{helsel2005}.  The log-normal distribution is often used to model left-censored data in environmental science applications \citep[e.g.,]{Krish2011}. The log-normal model density is 
\[ p_\theta(y) = \frac{1}{(2\pi \theta_2)^{1/2} y} \exp\Bigl\{-\frac{(\log y-\theta_1)^2}{2\theta_2} \Bigr\}, \]
where $\theta = (\theta_1, \theta_2)$ represent the mean and variance parameters for $\log Y$.  Again, log-normal is only the model for the observed concentrations---no model is assumed for the censored observations.  Figure~\ref{fig:censored}(a) displays the nonparametric estimator $\widehat G$ of the censored data distribution obtained by applying the Kaplan--Meier estimator with the censoring labels swapped: $t_i \mapsto 1-t_i$.  This $\widehat G$ is then used to define what we are referring to here as the ``exact'' IM contour via \eqref{eq:censored.boot.pl}, and then the corresponding Gaussian variational approximation---applied to $(\theta_1, \log\theta_2)$ first and then mapped back to $(\theta_1,\theta_2)$---is shown in Figure~\ref{fig:censored}(b).  This plot is very similar to that shown in \citet[][Fig.~10]{imcens} based on naive Monte Carlo, but far less expensive computationally.  
\end{ex}

\section{Conclusion}
\label{S:discuss}

In a similar spirit to the variational approximations that are now widely used in Bayesian statistics, and building on recent ideas presented in \citet{calibrated.boostrap}, we develop here a strategy to approximate the possibilistic IM's contour function---or at least its $\alpha$-cuts/level sets for specified $\alpha$---using ordinary Monte Carlo sampling and stochastic approximation.  A potpourri of applications is presented, from simple textbook problems to (parametric, nonparametric, and semiparametric) problems involving nuisance parameters, and even one of relatively high dimension, to highlight the flexibility, accuracy, and overall applicability of our proposal.  

There are a number of limitations associated with what has been proposed here in this paper.  Naturally, these limitations lead to open questions and future research directions to be pursued. First, the more sophisticated and efficient of the approximations proposed here, the one in Section~\ref{S:beyond}, is designed specifically for Gaussian variational families.  This is not a serious practical limitation since the Gaussian distribution is sure to be a good approximation when $n$ is moderate to large \citep{imbvm.ext}.  However, there are surely other variational families for which the density level sets have a clean, more-or-less closed-form expression in terms of the model parameters. Identifying other suitable models that can be used in this efficient approximation will give users more flexibility and ultimately lead to better and more accurate approximations.  Second, the proposed approximations depend on specifying a choice of $\alpha$ at the outset, which means that we are effectively only approximating certain features of the possibilistic IM's contour.  Arguably, the holistic nature of IM-based inference suggests that a broad rather than very specific approximation is preferred, and finding a way to stitch together these $\alpha$-specific IM approximations is an important open problem. Inspired by recent developments in \citet{calibrated.boostrap}, we believe that the answer to this question is {\em Yes}, and these details will be reported elsewhere.  Third, the details presented here are specifically focused on the case where prior information about the unknown $\Theta$ is vacuous.  Recent efforts \citep[e.g.,][]{martin.partial2} have focused on incorporating incomplete or partial prior information into the possibilistic IM construction; something along these lines is sure to be necessary as we scale up to problems of higher dimension.  The only downside to the incorporation of partial prior information is that the evaluation of the IM contour is often even more complicated than in the vacuous-prior case considered here.  This additional complexity means that efficient numerical approximations are even more important in these partial-prior formulations and, fortunately, we expect that the proposal here will carry over more-or-less directly.  Finally, with the exception of Example~\ref{ex:lasso}, our focus here was on problems involving relatively low-dimensional unknowns.  As is always the case, there are challenges associated with scaling up the proposed approximation strategy.  At the moment, we do not know how these challenges will be overcome, but we are inspired by the latest progress presented here and confident that they can be overcome. 


\bibliographystyle{apalike}
\bibliography{mybib}

\end{document}